\documentclass[a4paper,11pt]{article}
\usepackage[margin=2.5cm]{geometry}
\usepackage{hyperref}
\usepackage{amsmath,amsfonts,amssymb,array,graphicx,color,comment}

\graphicspath{{fig/}}
\numberwithin{equation}{section}

\newcommand{\aver}[1]{\langle {#1} \rangle}
\newcommand{\aaver}[1]{\langle\!\langle {#1} \rangle\!\rangle}
\newcommand{\mathfig}[1]{\raisebox{-0.5\height}{#1}}
\newcommand{\smallarray}[2]{\begin{array}{c} \scriptstyle #1 \\[-4pt] \scriptstyle #2 \end{array}}
\newcommand{\tetra}[6]{\left[\begin{array}{cccccc} #1 & #2 & #3 \\ #4 & #5 & #6 \end{array}\right]}
\newcommand{\arxiv}[1]{\href{https://arxiv.org/abs/#1}{\textsf{arXiv:#1}}}

\newcommand{\nn}{\nonumber}
\newcommand{\wh}{\widehat}
\newcommand{\id}{\mathbf{1}}
\newcommand{\Cbb}{\mathbb{C}}

\newcommand{\Zbb}{\mathbb{Z}}
\newcommand{\eps}{\varepsilon}
\newcommand{\TL}{\mathsf{TL}}
\newcommand{\cd}{c^\dag}

\newcommand{\F}{\mathcal{F}}
\newcommand{\I}{\mathcal{I}}

\newcommand{\Ab}{\overline{A}}
\newcommand{\Bb}{\overline{B}}
\newcommand{\Cb}{\overline{C}}
\newcommand{\Fb}{\overline{F}}
\newcommand{\Kb}{\overline{K}}
\newcommand{\Lb}{\overline{L}}
\newcommand{\Yb}{\overline{Y}}
\newcommand{\bb}{\bar{b}}
\newcommand{\fb}{\bar{f}}
\newcommand{\hb}{\bar{h}}
\newcommand{\ub}{\bar{u}}
\newcommand{\vb}{\bar{v}}
\newcommand{\wb}{\bar{w}}

\newcommand{\yb}{\bar{y}}

\newcommand{\lambdab}{\bar\lambda}
\newcommand{\mub}{\bar\mu}
\newcommand{\nub}{\bar\nu}

\newcommand{\vtheta}{\vartheta}
\newcommand{\vthetab}{\bar\vartheta}

\title{On a class of affine Temperley-Lieb modules}

\author{Yacine Ikhlef \\ Sorbonne Universit\'e, CNRS \\ Laboratoire de Physique Th\'eorique et Hautes \'Energies \\ LPTHE, F-75005 Paris, France}

\date{\today}

\begin{document}

\maketitle

\begin{abstract}
  We prove that the class of all affine Temperley-Lieb ($\TL^a$) modules, equipped with the recently introduced ``diagrammatic'' tensor product, forms a quasitensor (i.e. monoidal braided) category. When the $\TL^a$ parameter $q$ is not a root of unity, we consider the infinite subcategory generated by the simple quotients of reducible standard modules, and we argue that it is a semi-simple ribbon category. We present some evidence that, under the fusion rules and the associativity maps, each simple module in this subcategory is equivalent to a pair of $U_q(sl_2)$ heighest-weight modules.
\end{abstract}

\section{Introduction}

Temperley-Lieb algebras \cite{TL71}--\cite{MartinSaleur93} are diagram algebras which encode the structure of many solvable two-dimensional statistical models. The diagrams of the \emph{ordinary} Temperley-Lieb algebra $\TL(N)$ are drawn on a rectangle, whereas those of the \emph{affine} Temperley-Lieb algebra $\TL^a(N)$ \cite{FanGreen97}--\cite{GL98} live on an annulus. Therefore, both structures are essential for the complete description of a lattice model. The case of the affine Temperley-Lieb algebra $\mathcal{A}_N=\TL^a(N)$ is significantly more difficult to study, because (i) $\mathcal{A}_N$ is infinite-dimensional and (ii) there is no tensor product $\mathcal{A}_M\times \mathcal{A}_N\to \mathcal{A}_{M+N}$ for affine diagrams.
\medskip

Consider the category $\cal M$ of $\TL^a$ modules, in the sense of \cite{GL98}, i.e. a $\TL^a$ module $A$ is defined as a sequence of vector spaces $A(0),A(1),A(2)\dots$ where $A(N)$ corresponds to a system with $N$ nodes, and equipped with the action of $\TL^a$ diagrams from $A(N)$ to $A(N')$ (see Sec.~\ref{sec:TL} for more details). The problem of defining a tensor product $\mathcal{M}\times \mathcal{M}\to \mathcal{M}$, also called a ``fusion product'', is crucial for statistical physics, because it plays the same role as the operator product expansion in Quantum Field Theory. This fusion problem has been approached from various viewpoints over the recent years \cite{PRZ06}--\cite{IMD26}. In \cite{IMD26}, a ``diagrammatic'' tensor product $\otimes$ was introduced, inspired by the algebraic rules for connectivity diagrams in a disc with two holes, and it was argued that this tensor product is well-suited to describe intermediary states in correlation functions of connectivity operators. 
\medskip

The results of \cite{PRZ06}--\cite{IMD26} offer some promising perspectives for the understanding of correlation functions, especially in random geometry, namely statistical models of non-local lattice objects such as non-intersecting loops, polymers, spin interfaces, or percolation clusters \cite{Nienhuis84}--\cite{RRJ26}.
\medskip

In the present work, we obtain two sets of results concerning the $\TL^a$ modules. These results aim at a systematic description of correlations in $\TL$-invariant models, in analogy with the Virasoro algebra and Conformal Field Theory \cite{MooreSieb89}, based on the fact that the $\TL^a$ modules, equipped with the diagrammatic tensor product $\otimes$, can be organised into a category satisfying the coherence (pentagon and triangle) and braiding axioms \cite{Chari95}.

\paragraph{Main results.}
First, we consider the category $\cal M$ of $\TL^a$ modules, equipped with the diagrammatic tensor product $\otimes$ of \cite{IMD26}.
The unit element is the vacuum module, whose space $V(N)$ is spanned by the non-intersecting pairings (called link states) of $N$ nodes on a circle.
For any $A,B,C\in \mathcal{M}$, the structure maps
\begin{equation} \label{eq:struct1}
  \begin{aligned}
    & \alpha_{ABC}:A\otimes (B\otimes C) \to (A\otimes B)\otimes C \,,
    &\qquad& \sigma_{AB}:A\otimes B\to B\otimes A \,, \\
    & \ell_A:V\otimes A\to A \,,
    && r_A:A\otimes V\to A \,, \\
  \end{aligned}
\end{equation}
are given by \cite{IMD26}
\begin{equation} \label{eq:struct2}
  \begin{aligned}
    & \alpha_{ABC}(u\otimes(v\otimes w)) = (u\otimes v)\otimes w \\
    & \ell_A(v_0\otimes u) = r_A(u\otimes v_0) = u \\
    & \sigma_{AB}(u\otimes v) = \Omega^M \cdot (\Omega^{-M}v \otimes u)
  \end{aligned}
\end{equation}
where $u\in A(L)$, $v\in B(M)$, $w\in C(N)$, and $v_0$ is the unique link state of $V(0)$.
Here, $\Omega$ denotes the generator of rotations in $\TL^a$.
The monodromy reads
\begin{equation}
  \sigma_{BA} \cdot \sigma_{AB} = \theta_{A\otimes B} \cdot (\theta_A^{-1}\otimes \theta_B^{-1}) \,,
\end{equation}
where the twist automorphism $\theta_A$ is given by the central element $\theta=\Omega^N$ on $A(N)$.
In the present work, using elementary diagram calculus, we show that (i) $\alpha,\ell,r$ obey the triangle and pentagon relations, (ii) $\alpha,\sigma$ obey the hexagon relations, and (iii) the structure maps (\ref{eq:struct1}--\ref{eq:struct2}) are natural in their arguments $A,B,C$, which proves that $\cal M$ is a quasitensor category, i.e. it is monoidal and braided (see \cite{Chari95}).
\medskip

Then, when the $\TL^a$ parameter $q$ is not a root of unity, we consider a certain subcategory $\mathcal{Q} \subset \cal M$. To be specific, as proven in \cite{GL98}, the simple $\TL^a$ modules are the simple quotients of standard modules $W_{k,x}$ -- a class of modules defined in terms of link states with $2k$ legs ($2k=0,1,2,\dots$) and a twist factor $x\in \Cbb^\times$. We define $\cal Q$ as the category of simple quotients of \emph{reducible} standard modules $W_{k,-q^r}$, where $k\in \Zbb_{\geq 0}/2$ and $r\in k+\Zbb$ is such that $|r|>|k|$. These modules are conveniently parameterised by two positive integers $\lambda=|r+k|$ and $\lambdab=|r-k|$, and we denote the simple quotient of $W_{k,-q^r}$ as $Q_{k,-q^r}=[\lambda,\lambdab]$. In each such module, the quotienting translates into a ``null-state'' relation, which we use to compute the decomposition of $A \otimes B$ into a direct sum of simple modules, for any simple $A,B\in \cal Q$. Note that this approach is analogous to the derivation of fusion rules of degenerate Virasoro modules in CFT \cite{BPZ84}.
We find the decomposition
\begin{equation} \label{eq:decomp0} 
  [\lambda,\lambdab]\otimes [\mu,\mub]
  \equiv \bigoplus_{\smallarray{\nu=|\lambda-\mu|+1}{\lambda+\mu+\nu\text{ odd}}}^{\lambda+\mu-1}
  \bigoplus_{\smallarray{\nub=|\lambdab-\mub|+1}{\lambdab+\mub+\nub\text{ odd}}}^{\lambdab+\mub-1} [\nu,\nub] \,.
\end{equation}
Here, we recognise the $sl_2$ or $U_q(sl_2)$ fusion rules, independently for each set of labels $(\lambda,\mu,\nu)$ and $(\lambdab,\mub,\nub)$.
\medskip

Furthermore, we show that any simple module $A\in \mathcal{Q}$ is its own left and right dual, i.e. we construct a pair of morphisms $\eta_A:V\to A\otimes A$ and $\eps_A:A\otimes A\to V$ such that\footnote{Here we omit the insertion of $\alpha_{AAA}$, $\ell_A$ and $r_A$ to lighten the notation.}
\begin{equation} \label{eq:rigid0}
  (\id_A \otimes \eps_A) \cdot (\eta_A\otimes \id_A)
  = (\eps_A\otimes \id_A)\cdot(\id_A \otimes \eta_A) = \id_A \,.
\end{equation}
The morphism $\eta_A$ is determined by the ``seed state'' $\eta_A(v_0)=\xi \in (A\otimes A)(0)$, corresponding to the term $[1,1]$ in the RHS of \eqref{eq:decomp0} for $[\lambda,\lambdab]=[\mu,\mub]$.
For $\eps_A$, we use the fact that the Gram bilinear form $\aver{\,,\,}$ on $V\times V$ is non-degenerate \cite{GL98,DiF96}, and we define
\begin{equation}
  \eps_A(u) = \# \, \sum_i \aaver{u,\vthetab_i} \, \vtheta_i
\end{equation}
for any $u \in (A\otimes A)(N)$, where $\#$ stands for a normalisation constant, $(\vtheta_i),(\vthetab_i)$ is any pair of bases of $V(N)$ such that $\aver{\vtheta_i,\vthetab_j}=\delta_{ij}$, and $\aaver{\,,\,}$ is a uniquely defined invariant bilinear form on $(A\otimes A) \times V$.
The quantum dimension $d_A$ is then given by $\eps_A\cdot \eta_A = d_A \, \id_V$, and we find the expression in terms of $q$-integers:
\begin{equation}
  d_A = [\lambda]_{-q} \, [\lambdab]_{-q}
  \qquad \text{for } A=[\lambda,\lambdab] \,.
\end{equation}
Moreover, the subcategory $\cal Q$ equipped with the twist $\theta=\Omega^N$ has the structure of a ribbon category (see \cite{Chari95}).
\medskip

Finally, for any simple modules $A,B,C \in\mathcal{Q}$ such that $C$ appears in the decomposition of $A\otimes B$, we construct the ``splitting'' and ``fusing'' morphisms $\psi^C_{AB}:C\to A\otimes B$ and $\psi_C^{AB}:A\otimes B \to C$, in a way similar to $\eta_A$ and $\eps_A$, by introducing an invariant bilinear form $\aaver{\,,\,}$ on $(A\otimes B) \times \Cb$. Recall that, in a semi-simple monoidal category, the change of bases from $\mathrm{Hom}(D,A\otimes(B\otimes C))$ to $\mathrm{Hom}(D,(A\otimes B)\otimes C)$ is encoded by a matrix  $\F^D_{ABC}$, called the associativity matrix.
We argue that, in a well-chosen gauge (i.e. a particular normalisation of the morphisms $\psi_{AB}^C$ and $\psi_C^{AB}$), this matrix takes the form
\begin{equation}
  \F_{[a\bar a][b\bar b][c\bar c]}^{[d,\bar d]} =
  F_{abc}^d\otimes F_{\bar{a}\bar{b}\bar{c}}^{\bar{d}} \,,
\end{equation}
where $F_{abc}^d$ is the associativity matrix of $U_q(sl_2)$ heighest-weight modules, and here $\otimes$ denotes the usual tensor product of matrices.

\paragraph{Outline.}
The paper is organised as follows. In Sec.~\ref{sec:quasi}, we recall some definitions and the main results of \cite{IMD26}, and we prove that $\cal M$ is a quasitensor category. In Sec.~\ref{sec:Q}, we review the properties of standard modules, and we derive the decomposition formula~\eqref{eq:decomp0}. In Sec.~\ref{sec:Q2}, we show that $\cal Q$ is rigid, i.e. we construct the maps $\eta_A$ and $\eps_A$ satisfying \eqref{eq:rigid0} with the help of the Gram bilinear form, and we calculate the quantum dimension $d_A$. Using similar ideas, we construct the splitting and fusing maps $\psi_{AB}^C$ and $\psi^{AB}_C$, we study the braiding properties of $\cal Q$, and we calculate the associativity matrices $\F_{ABC}^D$ in some example cases. In Sec.~\ref{sec:conclusion}, we give some concluding remarks. In the Appendix, we gather some related results on irreducible standard modules, some technical details on the bilinear form $\aaver{\,,\,}$, and some formulas for the $6j$-symbols of $U_q(sl_2)$.

\section{The quasitensor structure of affine Temperley-Lieb modules}
\label{sec:quasi}

In this section, we briefly recall the definition of the affine Temperley-Lieb category $\TL^a$ with loop weight $\beta=-q-q^{-1}$ and its modules \cite{GL98}, and the construction of the diagrammatic tensor product from \cite{IMD26}. We then prove that the class of all $\TL^a$ modules forms a quasitensor category~$\cal M$, i.e. the associativity and commutativity isomorphisms are natural, and satisfy the coherence conditions and the hexagon relations.
Finally, we discuss some basic properties of invariant bilinear forms on $\TL^a$ modules, which will be useful in the next section.
These results are valid for any $q\in\Cbb^\times$.

\subsection{The affine Temperley-Lieb category and its modules}
\label{sec:TL}

\paragraph{The affine Temperley-Lieb category $\TL^a$.}
The affine Temperley-Lieb algebra $\TL^a(N)$ is the algebra generated by $e_0,e_1,\dots,e_{N-1}$ and $\Omega,\Omega^\dag$, subject to the relations
\begin{equation}
  \begin{aligned}
    & e_j^2 = \beta \, e_j \,,
    &\quad& e_je_{j\pm 1} e_j = e_j \,,
    &\quad& e_je_k=e_ke_j \quad \text{if }|j-k|>1 \\
    & \Omega\Omega^\dag = \Omega^\dag\Omega = \id \,,
    && e_j\Omega = \Omega e_{j+1} \,,
    && \Omega^2 e_1 = e_{N-1}e_{N-2}\dots e_1 \,,
  \end{aligned}
\end{equation}
where the indices $j,k$ are considered modulo $N$.
The loop weight is parameterised as $\beta=-q-q^{-1}$ with $q\in \Cbb^\times$.
The generators $e_j$ and $\Omega$ correspond to the diagrams
\begin{equation}
  e_j = \mathfig{
    \begin{picture}(0,0)%
      \includegraphics{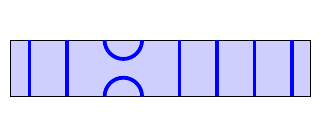}%
    \end{picture}%
    \setlength{\unitlength}{3947sp}%
    \begin{picture}(2574,1074)(3514,-3073)
      \put(4351,-2986){\makebox(0,0)[b]{\smash{\fontsize{11}{13.2}\normalfont {\color[rgb]{0,0,0}$j$}%
      }}}
      \put(5851,-2986){\makebox(0,0)[b]{\smash{\fontsize{11}{13.2}\normalfont {\color[rgb]{0,0,0}$N$}%
      }}}
      \put(3751,-2986){\makebox(0,0)[b]{\smash{\fontsize{11}{13.2}\normalfont {\color[rgb]{0,0,0}$1$}%
      }}}
      \put(4051,-2986){\makebox(0,0)[b]{\smash{\fontsize{11}{13.2}\normalfont {\color[rgb]{0,0,0}$\dots$}%
      }}}
      \put(5251,-2986){\makebox(0,0)[b]{\smash{\fontsize{11}{13.2}\normalfont {\color[rgb]{0,0,0}$\dots$}%
      }}}
      \put(4651,-2986){\makebox(0,0)[b]{\smash{\fontsize{11}{13.2}\normalfont {\color[rgb]{0,0,0}$j\!+\!1$}%
      }}}
    \end{picture}%
  }
  \qquad
  \Omega = \mathfig{
    \begin{picture}(0,0)%
      \includegraphics{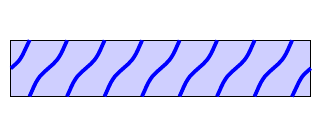}%
    \end{picture}%
    \setlength{\unitlength}{3947sp}%
    \begin{picture}(2574,1074)(3514,-3073)
      \put(5851,-2986){\makebox(0,0)[b]{\smash{\fontsize{11}{13.2}\normalfont {\color[rgb]{0,0,0}$N$}%
      }}}
      \put(3751,-2986){\makebox(0,0)[b]{\smash{\fontsize{11}{13.2}\normalfont {\color[rgb]{0,0,0}$1$}%
      }}}
      \put(4801,-2986){\makebox(0,0)[b]{\smash{\fontsize{11}{13.2}\normalfont {\color[rgb]{0,0,0}$\dots$}%
      }}}
    \end{picture}%
  }
\end{equation}
where, by convention, the diagrams act from top to bottom.
The above definition is only valid for $N\geq 2$.
For $N=1$, the algebra $\TL^a(1)$ is generated by $\Omega,\Omega^\dag$ subject to $\Omega\Omega^\dag = \Omega^\dag\Omega = \id$. For $N=0$, one defines $\TL^a(0)=\mathrm{span}(f^n, n\in \Zbb_{\geq 0})$.
\medskip

Following \cite{GL98}, one defines the affine Temperley-Lieb category $\TL^a$ as follows. The objects of $\TL^a$ are the non-negative integers $N=0,1,2,3\dots$, and the morphisms $N\to N'$ are the elements of $\TL^a(N',N)$, namely the linear combinations of non-intersecting diagrams on the annulus, which realise the pairing of $N$ incoming and $N'$ outgoing nodes.
The even and odd integers respectively form subcategories of $\TL^a$. One recovers the affine Temperley-Lieb algebra by setting $N=N'$, i.e. $\TL^a(N)=\TL^a(N,N)$.
\medskip

In \cite{IMD26} it was shown that the $\TL^a$ category is generated by the elementary diagrams $c_j\in\TL^a(N-2,N)$ and $\cd_j\in\TL^a(N,N-2)$, with $j=0,\dots,N-1$, subject to the relations
\begin{equation} \label{eq:TL.relations.cj}
  \begin{aligned}
    &\begin{cases}
      c_j^{\phantom\dag}\, c_{j+1}^\dag = c_{j+1}^{\phantom\dag} \, c_j^\dag= \id  & \text{if } j \geq 1 \\ 
      c_j^{\phantom\dag}\, c_j^\dag = \beta \, \id  & \text{for any } j 
    \end{cases}
    && \begin{cases}
      c_j c_k = c_k c_{j+2} \\
      c^\dag_k c^\dag_j = c^\dag_{j+2} c^\dag_k \\
      c^\dag_j c_k^{\phantom\dag} = c_k^{\phantom\dag} c^\dag_{j+2}
    \end{cases} \quad \text{if } 1\leq k \leq j \\
    & \begin{cases}
      c_0c_j =c_{j-1}c_0   \\
      c_0^{\phantom\dag}c^\dag_j =c^\dag_{j-1}c_0^{\phantom\dag} \\
      c_j^{\phantom\dag}c_0^\dag = c_0^\dag c_{j-1}^{\phantom\dag} \\
      c^\dag_jc_0^\dag = c_0^\dag c^\dag_{j-1}
      \end{cases} \quad \text{if } 2 \leq j\leq N-2
    &\quad& \begin{cases}
      c_0c_{N-1}^\dag = c_1 c_0^\dag \\
      c_{N-1}^{\phantom\dag}c_0^\dag = c_0^{\phantom\dag}c_1^\dag \\
      c_0^{\phantom\dag}c_1^\dag \cdot c_1^{\phantom\dag} c_0^\dag = c_1^{\phantom\dag} c_0^\dag \cdot c_0^{\phantom\dag}c_1^\dag = \id 
    \end{cases}
  \end{aligned}
\end{equation}
where by convention, the rightmost $c_i$ (resp. $c_i^\dag$) in these expressions is in $\TL^a(N-2,N)$ [resp. $\TL^a(N,N-2)$].
The generators $c_j$ and $\cd_j$ correspond to the diagrams
\begin{equation}
  c_j=\mathfig{
    \begin{picture}(0,0)%
      \includegraphics{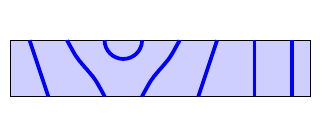}%
    \end{picture}%
    \setlength{\unitlength}{3947sp}%
    \begin{picture}(2574,1074)(3514,-3823)
      \put(3751,-2986){\makebox(0,0)[b]{\smash{\fontsize{11}{13.2}\normalfont {\color[rgb]{0,0,0}$1$}%
      }}}
      \put(4351,-2986){\makebox(0,0)[b]{\smash{\fontsize{11}{13.2}\normalfont {\color[rgb]{0,0,0}$j$}%
      }}}
      \put(5851,-2986){\makebox(0,0)[b]{\smash{\fontsize{11}{13.2}\normalfont {\color[rgb]{0,0,0}$N$}%
      }}}
      \put(4651,-2986){\makebox(0,0)[b]{\smash{\fontsize{11}{13.2}\normalfont {\color[rgb]{0,0,0}$j\!+\!1$}%
      }}}
      \put(4051,-2986){\makebox(0,0)[b]{\smash{\fontsize{11}{13.2}\normalfont {\color[rgb]{0,0,0}$\dots$}%
      }}}
      \put(5326,-2986){\makebox(0,0)[b]{\smash{\fontsize{11}{13.2}\normalfont {\color[rgb]{0,0,0}$\dots$}%
      }}}
\end{picture}%
  }
  \qquad \cd_j=\mathfig{
    \begin{picture}(0,0)%
      \includegraphics{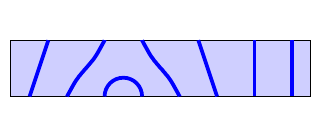}%
    \end{picture}%
    \setlength{\unitlength}{3947sp}%
    \begin{picture}(2574,1074)(3514,-3073)
      \put(3751,-2986){\makebox(0,0)[b]{\smash{\fontsize{11}{13.2}\normalfont {\color[rgb]{0,0,0}$1$}%
      }}}
      \put(4351,-2986){\makebox(0,0)[b]{\smash{\fontsize{11}{13.2}\normalfont {\color[rgb]{0,0,0}$j$}%
      }}}
      \put(5851,-2986){\makebox(0,0)[b]{\smash{\fontsize{11}{13.2}\normalfont {\color[rgb]{0,0,0}$N$}%
      }}}
      \put(4651,-2986){\makebox(0,0)[b]{\smash{\fontsize{11}{13.2}\normalfont {\color[rgb]{0,0,0}$j\!+\!1$}%
      }}}
      \put(5326,-2986){\makebox(0,0)[b]{\smash{\fontsize{11}{13.2}\normalfont {\color[rgb]{0,0,0}$\dots$}%
      }}}
      \put(4051,-2986){\makebox(0,0)[b]{\smash{\fontsize{11}{13.2}\normalfont {\color[rgb]{0,0,0}$\dots$}%
      }}}
    \end{picture}%
  }
\end{equation}
\medskip

The transposition $\lambda \mapsto \lambda^t$
(resp. conjugation $\lambda \mapsto \lambda^\dag$)
is defined as the linear
(resp. antilinear)
map from $\TL^a(N,N')$ to $\TL^a(N',N)$, which exchanges the incoming and outgoing nodes of every diagram.
\medskip

The ordinary Temperley-Lieb category $\TL$ is the subcategory of $\TL^a$, generated by the diagrams $c_j\in\TL^a(N-2,N)$ and $\cd_j\in\TL^a(N,N-2)$, with $j=1,\dots,N-1$ (the difference with $\TL^a$ is that $c_0,\cd_0$ are not included). The category $\TL$ admits a tensor product, defined by juxtaposing the diagrams side by side. If $\lambda\in\TL(L',L)$ and $\mu\in \TL(M',M)$, then $\lambda\otimes\mu\in\TL(L'+M',L+M)$. The $\TL$ category equipped with this tensor product is monoidal and braided~\cite{BSA18}.
\medskip

A \textit{central element} of $\TL^a$ is a collection $X=(X_0,X_1,X_2,\dots)$, where $X_N\in\TL^a(N)$, such that $X_{N'}\,\lambda=\lambda\,X_N$, for any $\lambda\in\TL^a(N',N)$. The full rotation $\theta_N=\Omega^N$, also called the twist operator, is a central element.
Consider the braid operators
\begin{equation} \label{eq:bj}
  b_j = q^{1/2}\id + q^{-1/2}e_j \,,
  \qquad \bb_j = q^{-1/2}\id + q^{1/2}e_j \,.
\end{equation}
The braid transfer matrices
\begin{equation} \label{eq:FN}
  F_N= c_0\cdot b_1b_2\dots b_N\cdot c_{N+1}^\dag \,,
  \qquad \Fb_N=c_0\cdot \bb_1\bb_2\dots \bb_N\cdot c_{N+1}^\dag 
\end{equation}
are also central elements.
\medskip

At $N=0$, the above operators reduce to
\begin{equation}
  \theta_0=\id \,,
  \qquad F_0=\Fb_0=c_1 \cd_0 = c_0 \cd_1 = f
\end{equation}
where $f$ is the single generator of $\TL^a(0)$.

\paragraph{The category $\cal M$ of $\TL^a$ modules.}
A \textit{$\TL^a$ module} is defined as a functor $A$ from $\TL^a$ to the category $\mathsf{Vect}_{\Cbb}$ of complex vector spaces:
$$ A: \begin{cases} N \mapsto A(N) \in \mathsf{Vect}_{\Cbb} \\ \lambda \mapsto A(\lambda) \,. \end{cases} \qquad N=0,1,2,\dots \qquad \lambda \in \TL^a(N',N) \,,$$
where $A(\lambda)$ is a linear map from $A(N)$ to $A(N')$.
The fact that $A$ is a functor means that it respects the identity morphisms and the composition of morphisms of $\TL^a$:
\begin{equation}
  A(\id_N) = \id_{A(N)} 
  \qquad A(\lambda'\lambda) = A(\lambda')\, A(\lambda) \,.
\end{equation}
The linear map $A(\lambda)$ is often simply denoted $\lambda$.
\medskip

Let $A$ and $B$ be two $\TL^a$ modules. A \textit{morphism} from $A$ to $B$ is defined as a natural transformation $\phi$ from $A$ to $B$. Concretely, it consists in a collection $\phi=(\phi_0,\phi_1,\phi_2,\dots)$, where each $\phi_N$ is a linear map from $A(N)$ to $B(N)$, and such that $\phi_{N'}\circ \lambda = \lambda \circ \phi_N$ for any diagram $\lambda \in \TL^a(N',N)$. We denote by $\mathrm{Hom}(A,B)$ the space of morphisms from $A$ to $B$.
We call $\cal M$ the category whose objects are the $\TL^a$ modules, and whose morphisms are defined this way.
\medskip

For instance, any central element $X$ yields a morphism $X_A:A \to A$ for every $\TL^a$ module $A$. Moreover, this morphism is natural in $A$, i.e. for any morphism $\phi:A \to B$, one has $X_B \circ \phi = \phi \circ X_A$.
\medskip

An invertible morphism is called an isomorphism.
For any $A,B\in \cal M$, we write $A\equiv B$ iff there exists an isomorphism between $A$ and $B$.

\paragraph{The bifunctor $\otimes:\mathcal{M}\times\mathcal{M}\to \mathcal{M}$.}
Let $A,B \in \mathcal{M}$.
We define the ``fused module'' $A\otimes B$ as~\cite{IMD26}
\begin{equation}
  (A\otimes B)(N) := \sum_{L,M\geq 0} \TL^a(N,L+M)\cdot
  \big(A(L)\otimes_{\Cbb} B(M) \big) \,,
\end{equation}
subject to the relations
\begin{subequations} \label{eq:fusion.rel}
  \begin{alignat}{1}
    & (\lambda\otimes \mu) \cdot (u\otimes v)
    = \lambda\,u \otimes \mu\,v \,,  \label{eq:fusion.rel.a} \\ 
    & \Omega \cdot (u\otimes v) = c_L\cdot (\Omega\,u \otimes c_0^\dag\,v) 
    \qquad \text{if } L\geq 1 \,,  \label{eq:fusion.rel.b}
  \end{alignat}
\end{subequations}
for any $u\in A(L)$, $v\in B(M)$, $\lambda\in\TL(L',L)$ and $\mu\in\TL(M',M)$. Here, $\otimes_\Cbb$ denotes the usual tensor product of vector spaces. In \cite{IMD26}, it is argued that the above relations reflect the algebraic rules for connectivity diagrams in a disc with two holes, where the nodes for $A$ and $B$ are located on the boundary of the holes, and those for $A\otimes B$ live on the external boundary of the disc.
\medskip

Consider a collection of linear maps $\phi_{L,M}:A(L)\otimes_\Cbb B(M) \to C(L+M)$ which preserve the relations \eqref{eq:fusion.rel}, i.e.
\begin{equation} \label{eq:phi.preserves.rel}
  \begin{aligned}
    & (\lambda\otimes \mu) \cdot \phi_{L,M}(u\otimes v)
    = \phi_{L',M'}(\lambda\,u \otimes \mu\,v) \,, \\ 
    & \Omega \cdot \phi_{L,M}(u\otimes v)
    = c_L\cdot \phi_{L,M+2}(\Omega\,u \otimes c_0^\dag\,v)
    \qquad \text{if } L\geq 1 \,,
  \end{aligned}
\end{equation}
for any $u\in A(L)$, $v\in B(M)$, $\lambda\in\TL(L',L)$ and $\mu\in\TL(M',M)$.
Then the linear map
\begin{equation}
  \phi_N\big(\lambda\cdot(u\otimes v)\big):=\lambda \cdot \phi_{L,M}(u\otimes v)
\end{equation}
for $u\in A(L)$, $v\in B(M)$ and $\lambda\in\TL^a(N,L+M)$ is well-defined on $(A\otimes B)(N)$, and $\phi=(\phi_0,\phi_1,\phi_2,\dots)$ is a morphism from $(A\otimes B)$ to $C$.
\medskip

In particular, for any pair of morphisms $\phi:A\to A'$ and $\psi:B \to B'$, the collection of linear maps
\begin{equation}
  \left\{ \begin{array}{ccc}
    A(L) \otimes_\Cbb B(M) &\to& (A'\otimes B')(L+M) \\
    u\otimes v &\mapsto& \phi(u) \otimes \psi(v)
  \end{array} \right.
\end{equation}
obeys \eqref{eq:phi.preserves.rel}, and hence it yields a morphism from $(A\otimes B)$ to $(A' \otimes B')$, which we denote $(\phi\otimes \psi)$.
It is easy to verify that 
\begin{equation}
  \id_A\otimes \id_B = \id_{A\otimes B} \,,
  \qquad
  (\phi'\circ\phi) \otimes (\psi'\circ\psi)
  = (\phi'\otimes\psi') \circ (\phi\otimes \psi) \,.
\end{equation}
Thus, the maps $(A,B)\mapsto A \otimes B$ and $(\phi,\psi)\mapsto \phi \otimes \psi$  define a bifunctor on the category $\cal M$, which we simply call the tensor product $\otimes$.

\paragraph{Symmetries.}
The reflection $\rho$ is defined as the linear map in $\TL^a(N,N')$, which exchanges respectively the incoming nodes $j \leftrightarrow N'+1-j$ and the outgoing nodes $j \leftrightarrow N+1-j$. In $\TL^a(N)$, we then have $\rho(e_j)=e_{N-j}$ and $\rho(\Omega)=\Omega^\dag$, and $\rho$ is an automorphism of the algebra $\TL^a(N)$.
Similarly, we define the sign automorphism $\tau$ by its action on the elementary diagrams
\begin{equation} \label{eq:def.tau}
  \tau(c_j)=(-1)^{\delta_{j0}}c_j \,, \qquad \tau(\cd_j)=(-1)^{\delta_{j0}}\cd_j \,.
\end{equation}
For any $A\in\cal M$, we denote by $\rho(A)$ and $\tau(A)$, respectively, the modules induced by $\rho$ and $\tau$ on $A$.
It is then easy to show that the following isomorphisms hold:
\begin{equation} \label{eq:sym.fusion}
  \rho(A\otimes B) \equiv \rho(A) \otimes \rho(B) \,,
  \qquad \tau(A\otimes B) \equiv  \tau(A)\otimes B
  \equiv A\otimes \tau(B) 
\end{equation}
for any $A,B\in \cal M$.

\subsection{The quasitensor structure}

\paragraph{Associativity.}
The associativity of the tensor product $\otimes$ is based on the following observation.
For any $A,B\in \cal M$, one has the identities in $A\otimes B$:
\begin{equation}
  \begin{aligned}
    & u \otimes c_j\,v = c_{L+j}\cdot (u\otimes v) \,,
    &\qquad& 1 \leq j\leq M-1 \,, \\
    & u \otimes \cd_j\,v = \cd_{L+j}\cdot (u\otimes v) \,,
    && 1 \leq j\leq M+1 \,, \\
    & u \otimes c_0\,v = c_0\,c_{L+2}\cdot (\cd_0\,u \otimes v) \,, \\
    & u \otimes \cd_0\,v = \Omega \cdot (\cd_0\,u \otimes v) \,,
  \end{aligned}
\end{equation}
for any $u\in A(L)$ and $v\in B(M)$.
These relations can be written as
\begin{equation}
  u \otimes c_j\,v = a(c_j) \cdot \left(c_0^{\dag m(c_j)}u \otimes v \right) \,,
  \qquad u \otimes \cd_j\,v = a(\cd_j) \cdot \left(c_0^{\dag m(\cd_j)}u \otimes v \right) \,,
\end{equation}
where $a(c_j),a(\cd_j)$ are words on the alphabet $\mathcal{A}=\{c_0,\cd_0,c_1,\cd_1,\dots\}$ (recall that $\Omega=c_1\cd_0$), and $m(c_j),m(\cd_j)$ are non-negative integers.
Thus, for any word $\lambda=\lambda_1\lambda_2\dots \lambda_n$ where $\lambda_i\in\mathcal{A}$, we define
\begin{equation}
  a(\lambda):=a(\lambda_1)\dots a(\lambda_n) \,,
  \qquad m(\lambda):=m(\lambda_1)+\dots+m(\lambda_n) \,,
\end{equation}
and we have
\begin{equation} \label{eq:u.otimes.lam.v}
  u\otimes \lambda\,v = a(\lambda)\cdot \left(c_0^{\dag m(\lambda)} u\otimes v\right) \,.
\end{equation}
Now, let $A,B,C \in \cal M$, and consider the map
\begin{equation} \label{eq:assoc-map}
  \alpha_{ABC}:u\otimes \big(\lambda\cdot (v\otimes w) \big) \mapsto
  a(\lambda) \cdot \Big( \big((\cd_0)^{m(\lambda)}u \otimes v \big) \otimes w \Big) \,,
\end{equation}
for $u\in A(L)$, $v\in B(M)$, $w\in C(K)$.
In \cite{IMD26}, it is shown that (i) the map $\alpha_{ABC}$ is well-defined for any diagram $\lambda\in \TL^a(J,M+K)$, i.e. it preserves the relations \eqref{eq:TL.relations.cj}, and (ii) it obeys \eqref{eq:phi.preserves.rel} for the two consecutive tensor products.
Thus \eqref{eq:assoc-map} yields a morphism from $A\otimes(B\otimes C)$ to $(A\otimes B)\otimes C$, called the associator.
\medskip

Similarly, for any word $\lambda$ on $\cal A$, one can construct a word $\wh{a}(\lambda)$ and an integer $\wh{m}(\lambda)$ such that, for any $u \in A(L)$ and $v \in B(M)$
\begin{equation} \label{eq:lam.u.otimes.v}
  \lambda\, u\otimes v = \wh{a}(\lambda)\cdot (u\otimes c_0^{\dag \wh{m}(\lambda)} v) \,.
\end{equation}
With an argument similar to the above, one shows that the map
\begin{equation}
  \wh\alpha_{ABC}:\big(\lambda\cdot (u\otimes v) \big) \otimes w\mapsto
  \wh{a}(\lambda) \cdot \Big( u\otimes \big(v \otimes (\cd_0)^{\wh{m}(\lambda)}w \big) \Big) 
\end{equation}
yields a morphism from $(A\otimes B)\otimes C$ to $A\otimes (B\otimes C)$.
\medskip

As a consequence of \eqref{eq:u.otimes.lam.v} and \eqref{eq:lam.u.otimes.v}, the space $(A\otimes (B\otimes C))(N)$ [resp. $((A\otimes B)\otimes C)(N)$] is spanned by elements of the form $\lambda\cdot (u\otimes (v\otimes w))$ [resp. $\lambda\cdot ((u\otimes v)\otimes w)$], with $u\in A(L)$, $v\in B(M)$, $w\in C(K)$ and $\lambda\in \TL^a(N,L+M+K)$. For these elements, we have
\begin{equation}
  \begin{aligned}
    &\alpha_{ABC}[\lambda\cdot (u\otimes (v\otimes w))] = \lambda\cdot ((u\otimes v)\otimes w) \,, \\
    &\wh\alpha_{ABC}[\lambda\cdot ((u\otimes v)\otimes w)] = \lambda\cdot (u\otimes (v\otimes w)) \,,
  \end{aligned}
\end{equation}
and thus the morphisms $\alpha_{ABC}$ and $\wh\alpha_{ABC}$ are inverse of one another.
Therefore, we have the isomorphism
\begin{equation}
  A\otimes (B\otimes C) \equiv (A\otimes B) \otimes C
\end{equation}
for any $A,B,C \in \cal M$.
\medskip

Moreover, for any $\TL^a$ modules $A,B,C$ and $A',B',C'$ and any morphisms $\phi:A\to A'$, $\psi:B\to B'$ and $\eta:C\to C'$, we have
\begin{alignat}{1}
  ((\phi\otimes \psi)\otimes \eta) [\alpha_{ABC}(u\otimes(v\otimes w))]
  &= (\phi(u)\otimes \psi(v)) \otimes \eta(w) \nn \\
  &= \alpha_{A'B'C'} [(\phi \otimes (\psi\otimes \eta))(u\otimes(v\otimes w))] \,,
\end{alignat}
and thus
\begin{equation}
  ((\phi\otimes \psi)\otimes \eta) \circ \alpha_{ABC}
  = \alpha_{A'B'C'} \circ (\phi \otimes (\psi\otimes \eta)) \,,
\end{equation}
which proves that $\alpha_{ABC}$ is natural with respect to its arguments $A,B,C$.

\paragraph{Unit element.}
The \textit{vacuum space} $V(N)$ is defined as the linear span of the non-intersecting pairings of $N$ nodes living on the boundary of a semi-infinite cylinder (or, equivalently, a disc). The spaces $V(N)$ with the diagrammatic action of $\TL^a(N',N)$ form a $\TL^a$ module (see \cite{IMD26}). We denote by $v_0$ the single basis element of $V(0)$. Then any $v\in V(N)$ can be written uniquely as $v=\lambda\cdot v_0$, with $\lambda\in\TL(N,0)$. For any $A\in \cal M$, consider the maps
\begin{equation}
  \ell_A: \left\{\begin{array}{ccc}
  (V \otimes A)(M+N) &\to& A(M+N) \\
  \lambda\,v_0 \otimes u &\mapsto& (\lambda\otimes \id_N)\cdot u
  \end{array}\right.
  \qquad \wh{\ell}_A(u): \left\{\begin{array}{ccc}
   A(N)&\to& (V \otimes A)(N) \\
   u &\mapsto& v_0\otimes u
  \end{array}\right.
\end{equation}
with $\lambda\in \TL(M,0)$ and $u\in A(N)$.
The map $\ell_A$ obeys \eqref{eq:phi.preserves.rel}, and hence it is well-defined and yields a morphism of $\TL^a$ modules. The map $\wh{\ell}_A$ is also a morphism, and it is the inverse of $\ell_A$.
For any morphism $\phi:A\to A'$ we have
\begin{equation}
  \phi(\ell_A(\lambda v_0\otimes u)) = \phi((\lambda\otimes \id_N)\cdot u)
  = (\lambda\otimes \id_N)\cdot \phi(u) = \ell_{A'}(\lambda v_0\otimes\phi(u)) \,,
\end{equation}
for any $\lambda\in \TL(M,0)$ and $u\in A(N)$, and hence $\ell_A$ is a natural isomorphism from $V\otimes A$ to $A$, called the left unitor.
\medskip

Similarly, the map $r_A:u \otimes \lambda\,v_0 \mapsto (\id_N\otimes \lambda)\cdot u$ yields a natural isomorphism from $A\otimes V$ to $A$, called the right unitor.
\medskip

We thus have the natural isomorphisms, for any $\TL^a$ module $A$
\begin{equation}
  V\otimes A \equiv A\otimes V \equiv A \,,
\end{equation}
and therefore $V$ is the unit element for the tensor product $\otimes$.

\paragraph{Coherence conditions.}
Let us check that the structure maps $(\alpha,\ell,r)$ satisfy the following coherence conditions
\begin{equation}
  \begin{aligned}
    & (r_A \otimes \id_B)\circ \alpha_{AVB} = \id_A\otimes \ell_B \,, \\
    & \alpha_{A\otimes B,C,D}\circ \alpha_{A,B,C\otimes D}
    = (\alpha_{ABC} \otimes \id_D) \circ \alpha_{A,B\otimes C,D}
    \circ(\id_A \otimes \alpha_{BCD}) \,,
  \end{aligned}
\end{equation}
respectively called the triangle and pentagon identities.
\medskip

For the triangle identity, we write
\begin{alignat}{1}
  & \big((r_A \otimes \id_B)\circ \alpha_{AVB}\big)
  \big(u\otimes (\lambda\,v_0\otimes v) \big) \nn \\
  &\qquad= (r_A \otimes \id_B)
  \big( (u\otimes \lambda\,v_0)\otimes v \big)
  = \big((\id_L\otimes \lambda)u\big)\otimes v \nn \\
  &\qquad= (\id_L\otimes \lambda \otimes \id_M)
  \cdot(u\otimes v)
  = u\otimes \big((\lambda\otimes \id_M)\,v\big) \nn\\
  &\qquad= (\id_A\otimes \ell_B)\big(u\otimes(\lambda\,v_0\otimes v) \big) \,,
\end{alignat}
for any $u\in A(L)$, $v\in B(M)$, and $\lambda\in\TL(N,0)$.
The pentagon identity is easily verified on states of the form $(u\otimes(v\otimes(w\otimes x)))$.
\medskip

Therefore, the category $\cal M$ of $\TL^a$ modules, equipped with the tensor product $\otimes$, the unit element $V$, and the structure maps $(\alpha,\ell,r)$, forms a monoidal category.

\paragraph{Braiding and monodromy.}
For any $A,B \in \cal M$, the linear map defined as
\begin{equation} \label{eq:sigma}
  \sigma_{AB}(u\otimes v) := \Omega^M\cdot (\Omega^{-M}v\otimes u)
\end{equation}
for any $u\in A(L)$ and $v\in B(M)$, obeys \eqref{eq:phi.preserves.rel}, and thus it yields a morphism from $(A\otimes B)$ to $(B\otimes A)$. This morphism is invertible, and its inverse reads
\begin{equation}
  \sigma_{AB}^{-1}(v\otimes u) = \Omega^{-M}\cdot (u\otimes \Omega^{M}v) \,.
\end{equation}
Moreover, for any pair of morphisms $\phi:A\to A'$ and $\psi:B \to B'$, one easily checks that $(\psi\otimes \phi)\circ \sigma_{AB} = \sigma_{A'B'}\circ (\phi\otimes \psi)$. Hence, $\sigma_{AB}$ is a natural isomorphism from $(A\otimes B)$ to $(B\otimes A)$, called the commutor.
\medskip

Let us check that the structure maps $\alpha$ and $\sigma$ satisfy the hexagon relations
\begin{equation}
  \begin{aligned}
    & \alpha_{CAB} \circ \sigma_{A\otimes B,C} \circ \alpha_{ABC}
    = (\sigma_{AC} \otimes \id_B) \circ \alpha_{ACB}
    \circ (\id_A\otimes \sigma_{BC}) \,, \\
    & \alpha_{BCA}^{-1} \circ \sigma_{A,B\otimes C} \circ \alpha_{ABC}^{-1}
    = (\id_B \otimes \sigma_{AC}) \circ \alpha_{BAC}^{-1} \circ (\sigma_{A\otimes B}\otimes \id_C) \,.
  \end{aligned}
\end{equation}
Using \eqref{eq:fusion.rel.b} repeatedly, we obtain the identity
\begin{equation}
  \Omega^n\cdot(u\otimes v) = c_{L-n+1}\dots c_L\cdot(\Omega^n u \otimes c^{\dag n}_0 v) \,,
\end{equation}
for any $u\in A(L)$, $v\in B(M)$ and integer $n$ such that $0\leq n\leq L$.
In particular, this yields the alternative expression for $\sigma_{AB}$:
\begin{equation}
  \sigma_{AB}(u\otimes v) = c_1\dots c_M\cdot(v \otimes c^{\dag M}_0 u) \,.
\end{equation}
We then write, for any $u\in A(L)$, $v\in B(M)$ and $w\in C(N)$
\begin{alignat}{1}
  & (\alpha_{CAB} \circ \sigma_{A\otimes B,C} \circ \alpha_{ABC})
  (u\otimes(v\otimes w))
  = \Omega^N \cdot ((\Omega^{-N}w\otimes u)\otimes v) \nn\\
  &\qquad= (\sigma_{AC}\otimes \id_B) \, \Omega^{N} \cdot ((\Omega^{-N}(u\otimes w))\otimes v) \nn\\
  &\qquad= (\sigma_{AC}\otimes \id_B)\, c_{L+1}\dots c_{L+N} \cdot ((u\otimes w)\otimes c^{\dag N}_0  v) \nn\\
  &\qquad= ((\sigma_{AC}\otimes \id_B)\circ \alpha_{ACB}) \, c_{L+1}\dots c_{L+N} \cdot ( u\otimes (w \otimes c^{\dag N}_0  v)) \nn\\
  &\qquad= ((\sigma_{AC}\otimes \id_B)\circ \alpha_{ACB}) ( u\otimes (c_1\dots c_N (w \otimes c^{\dag N}_0  v))) \nn\\
  &\qquad= ((\sigma_{AC}\otimes \id_B)\circ \alpha_{ACB}\circ (\id_A\otimes \sigma_{BC})) (u\otimes (v\otimes w)) \,,
\end{alignat}
and
\begin{alignat}{1}
  & (\wh\alpha_{BCA} \circ \sigma_{A,B\otimes C} \circ \alpha^{-1}_{ABC})((u\otimes v)\otimes w)
  = c_1\dots c_{M+N} \cdot (v\otimes (w\otimes c^{\dag (M+N)}_0 u)) \nn \\
  &\qquad = c_1\dots c_M \cdot (v\otimes c_1\dots c_N(w\otimes c^{\dag (M+N)}_0 u)) \nn \\
  &\qquad = (\id_B \otimes \sigma_{AC}) \, c_1\dots c_M \cdot (v\otimes (c^{\dag M}_0 u\otimes w)) \nn \\
  &\qquad = ((\id_B \otimes \sigma_{AC})\circ \alpha^{-1}_{BAC}) \, c_1\dots c_M \cdot ((v\otimes c^{\dag M}_0 u)\otimes w) \nn \\
  &\qquad = ((\id_B \otimes \sigma_{AC})\circ \alpha^{-1}_{BAC}) \, ((c_1\dots c_M\,(v\otimes c^{\dag M}_0 u))\otimes w) \nn \\
  &\qquad = ((\id_B\otimes \sigma_{AC})\circ \alpha^{-1}_{BAC}\circ(\sigma_{AB}\otimes \id_C)) ((v\otimes u)\otimes w) \,.
\end{alignat}
Therefore, the monoidal category $\cal M$, equipped with the commutor $\sigma$, is braided. The braid (or Yang-Baxter) relation is automatically satisfied in a braided monoidal category. It reads
\begin{equation}
  (\sigma_{BC}\otimes \id_A) \circ (\id_B \otimes \sigma_{AC})
  \circ (\sigma_{AB}\otimes \id_C)
  = (\id_C\otimes \sigma_{AB}) \circ (\sigma_{AC}\otimes \id_B)
  \circ (\id_A \otimes \sigma_{BC}) \,,
\end{equation}
where we have dropped the associators.
\medskip

Finally, let us discuss the monodromy between two $\TL^a$ modules $A$ and $B$, which is defined as $G_{AB}:=\sigma_{BA}\circ\sigma_{AB}$. By construction, $G_{AB}$ is a natural automorphism of $A\otimes B$. It reads
\begin{equation}
  G_{AB}(u\otimes v) = (\sigma_{BA} \circ \sigma_{AB})(u\otimes v)
  = \Omega^{L+M} \cdot (\Omega^{-L}u\otimes \Omega^{-M}v) \,,
\end{equation}
for any $u\in A(L)$ and $v\in B(M)$.
This can be written as
\begin{equation} \label{eq:g.theta}
  G_{AB} = \theta_{A\otimes B} \circ (\theta_A^{-1} \otimes \theta_B^{-1}) \,,
\end{equation}
where $\theta_A:=\Omega^N$ on $A(N)$ defines a natural automorphism, called the twist.

\subsection{Invariant bilinear forms}
Let $A,B\in \cal M$, and $g=(g_0,g_1,g_2,\dots)$ be a collection of bilinear forms $g_N:A(N) \times B(N)\to \Cbb$. We say that $g$ is an invariant bilinear form on $A \times B$ iff it satisfies
\begin{equation}
  g_N(u, \lambda\,v) = g_M(\lambda^t\,u, v)
\end{equation}
for any $u\in A(N)$, $v \in B(M)$ and $\lambda \in \TL^a(N,M)$.
We denote by $\I(A,B)$ the space of invariant bilinear forms on $A \times B$.
Like for morphisms, to lighten the notation, we often write $g(u,v)$ instead of $g_N(u,v)$.
\medskip

For any $g\in \I(A,B)$, we define the left and right radicals of $g$ as
\begin{equation}
  \begin{aligned}
    &\overline{\rm Rad}(g,N):= \{u \in A(N) | \ \forall v\in B(N) \,, g(u,v)=0 \} \,, \\
    & \mathrm{Rad}(g,N):=\{v \in B(N) | \ \forall u\in A(N) \,, g(u,v)=0 \} \,.
  \end{aligned}
\end{equation}
Then $\overline{\rm Rad}(g)$ is a submodule of $A$, and $\mathrm{Rad}(g)$ is a submodule of $B$. We say that $g$ is non-degenerate iff $\overline{\rm Rad}(g)=0$ and $\mathrm{Rad}(g)=0$. If the pair $(A,B)$ admits a non-degenerate form $g\in\I(A,B)$, then for any $N$ one has $\dim A(N)=\dim B(N)$, and there exists a pair of bases $(x_i), (y_i)$ of $A(N)$ and $B(N)$ respectively, such that $g(x_i,y_j)=\delta_{ij}$. We say that $(x_i)$ and $(y_i)$ are mutually dual bases.
\medskip

Consider two pairs of $\TL^a$ modules $(A,B)$ and $(A',B')$, and two morphisms $\phi:A\to A'$ and $\psi:B \to B'$. For any $g \in \I(A',B')$, the bilinear map defined as
\begin{equation}
  h(u,v):=g[\phi(u),\psi(v)] \,,
\end{equation}
for any $u \in A(N)$ and $v \in B(N)$, is an element of $\I(A,B)$. Moreover, if $\phi$ and $\psi$ are invertible, then we have the isomorphism of vector spaces
\begin{equation} \label{eq:F=F'}
  \begin{array}{ccc}
    \I(A',B') & \longleftrightarrow & \I(A,B) \\
    g(u,v)=h[\phi^{-1}(u),\psi^{-1}(v)] && h(u,v)=g[\phi(u),\psi(v)]
  \end{array}
\end{equation}
\medskip

Let $A,B,B'$ be three $\TL^a$ modules, such that there exists a non-degenerate form  $\aver{\,,\,} \in \I(B,B')$. We then denote by $(y_i),(y'_i)$ a pair of mutually dual bases of $B(N)$ and $B'(N)$. We have the isomorphism of vector spaces
\begin{equation} \label{eq:Hom=F}
  \begin{array}{ccc}
    \mathrm{Hom}(A,B) & \longleftrightarrow & \I(A,B') \\
    \phi(u)=\sum_i g(u,y'_i)\, y_i && g(u,v)=\aver{\phi(u),v}
  \end{array}
\end{equation}
and, in this expression, the morphism $\phi$ is independent of the choice of mutually dual bases $(y_i),(y'_i)$.
\medskip

Let $A,B,C$ be three $\TL^a$ modules, and let us denote $\Ab=\rho(A)$ and $\Bb=\rho(B)$, where $\rho$ is the reflection automorphism.
Consider the linear maps
\begin{equation} \label{eq:Kr}
  K_r: \left\{\begin{array}{ccc}
  \I(A \otimes B,C) &\to& \I(A, C\otimes \Bb) \\
  g &\mapsto& K_r(g)(u,w\otimes \vb)=g[c_{N+1}\dots c_{N+M} (u\otimes \rho(\vb)), w]
  \end{array}\right.
\end{equation}
for any $u\in A(M+N)$, $\vb\in \Bb(M)$ and $w\in C(N)$, and
\begin{equation} \label{eq:Kl}
  K_\ell: \left\{\begin{array}{ccc}
  \I(A \otimes B,C) &\to& \I(B, \Ab\otimes C) \\
  g &\mapsto& K_\ell(g)(v,\ub \otimes w) = g[c_1\dots c_L (\rho(\ub)\otimes v), w]
  \end{array}\right.
\end{equation}
for any $\ub\in \Ab(L)$, $v\in B(L+N)$ and $w\in C(N)$.
Each of these maps define an isomorphism of vector spaces.

\section{The simple quotients of reducible standard modules}
\label{sec:Q}

In this section, we consider the category $\cal M$ of $\TL^a$ modules, in the case when $q$ is not a root of unity.
We introduce the infinite subcategory $\cal Q$ generated by the simple quotients of reducible standard modules.
Using the explicit form of equivalence relations (``null-state'' conditions) in this particular class of modules, we calculate their tensor products, and we find that they consist in finite direct sums of simple modules, proving that $\cal Q$ is semi-simple.

\subsection{Simple quotients of standard modules}

Let $k\in\Zbb_{\geq 0}/2$ and $x\in \Cbb^\times$.
Recall that the standard module $W_{k,x}$ \cite{GL98} is given by the collection of vector spaces
$$W_{k,x}(0), W_{k,x}(1), W_{k,x}(2) \dots$$
where (i) $W_{k,x}(N)=0$ if $N<2k$ or $N-2k$ is odd; (ii) otherwise, $W_{k,x}(N)$ is the span of link states of $N$ nodes on the boundary of the half-infinite cylinder, with $2k$ nodes attached by legs to a marked point, and $N-2k$ arches connecting the other nodes. For each link state, we draw a reference line connecting the marked point with the midpoint of nodes $N$ and $1$. The action of a diagram $\lambda\in \TL^a(N',N)$ on a link state $u\in W_{k,x}(N)$ is obtained by stacking $u$ above $\lambda$ and reading the resulting link state, with a factor $\beta$ for each contractible loop, and (i) a factor $(x+x^{-1})$ for each non-contractible loop if $k=0$ (ii) a factor $x$ (resp. $x^{-1}$) for each leg crossing the reference line with the marked point on its right (resp. on its left) if $k>0$.
The dimension of $W_{k,x}(N)$ is given by
\begin{equation} \label{eq:dk}
  \dim W_{k,x}(N) = d_k(N) = \begin{cases}
    \left(\begin{array}{cc} N \\ \frac N2-k \end{array}\right)
    & \text{if } N\in 2k+2\Zbb_{\geq 0} \\
    0 &\text{otherwise.}
  \end{cases}
\end{equation}
\medskip

We can write
\begin{equation}
  W_{k,x}(N) = \TL^a(N,2k) \cdot u_k \,,
\end{equation}
where $u_k$ is the unique link state of $W_{k,x}(2k)$, and satisfies
\begin{equation} \label{eq:seed}
  \begin{aligned}
    & c_0 \cdot u_k = 0 && \text{if } k\geq 1 \\
    & \Omega \cdot u_k =x\, u_k && \text{if } k\geq \frac12 \\
    & f\cdot u_k = (x+x^{-1})\, u_k && \text{if } k=0 \,.
  \end{aligned}
\end{equation}
More generally, for any $\TL^a$ module $A$, any nonzero element $\xi\in A(2k)$ which obeys the relations~\eqref{eq:seed} is called a seed state of type $(k,x)$. It is then easy to show that the linear map 
\begin{equation} \label{eq:insertion}
  \lambda\cdot u_k \mapsto \lambda\cdot \xi
\end{equation}
is well-defined for any $\lambda\in\TL^a(N,2k)$, and yields a morphism $W_{k,x}\to A$, which we call the insertion map associated to $\xi$. In the following, it will be convenient to define $W_{-k,x}=W_{k,1/x}$, and, correspondingly, to identify the types $(-k,x)\equiv (k,1/x)$.
\medskip

There exists a unique invariant bilinear form $\aver{\,,\,}$ on $W_{k,1/x} \times W_{k,x}$, called the Gram form, such that $\aver{\ub_k,u_k}=1$, where $\ub_k$ denotes the unique link state of $W_{k,1/x}$. It was proved in \cite{GL98} that (i) the maximal submodule of $W_{k,x}$ is the radical of the Gram form; (ii) $W_{k,x}$ is reducible iff $x$ is of the form $x=\pm q^r$ with $r\in k+\Zbb$ and $|r|>k$; (iii) in the latter case, $W_{k,\pm q^r}$ admits a seed state of type $(r,\pm q^k)$.
\medskip

Since we assume here that $q$ is not a root of unity, it is easy to deduce that, for any $k\in \Zbb_{\geq 0}/2$ and $r\in k+\Zbb$ such that $|r|>k$, the maximal submodule of $W_{k,\pm q^r}$ is isomorphic to the irreducible module $W_{r,\pm q^k}$. We then define the simple quotient
\begin{equation} \label{eq:Q=W/W}
  Q_{k,\pm q^r}:= W_{k,\pm q^r} \big/ W_{r,\pm q^k} \,.
\end{equation}
The dimensions of the corresponding vector spaces are given by
\begin{equation} \label{eq:dim.Q}
  \dim Q_{k,\pm q^r}(N) = d_k(N) - d_r(N) \,,
\end{equation}
where $d_k(N)$ is defined in \eqref{eq:dk}. The seed state for the radical of the Gram form in $W_{k,\pm q^r}$ can be written \cite{IMD24}
\begin{equation}
  \chi_{k,r} = P_{|2r|}\, c_0^{\dag(|r|-k)} \cdot u_k \,,
\end{equation}
where $P_m$ denotes the Jones-Wenzl projector of size $m$. It is then convenient to view $Q_{k,\pm q^r}$ as the quotient of $W_{k,\pm q^r}$ by a ``null-state'' condition
\begin{equation}
  Q_{k,\pm q^r}:= W_{k,\pm q^r} \big/ \aver{\chi_{k,r}=0} \,.
\end{equation}
Using this equivalence and the dimension formula \eqref{eq:dim.Q}, one easily shows that the link states with strictly less than $(|r|-k)$ arches crossing the reference line form a basis of $Q_{k,\pm q^r}(N)$.
\medskip

Since the modules $Q_{k,+q^r}$ and $Q_{k,-q^r}$ are related through the sign automorphism \eqref{eq:def.tau}, in the following, for simplicity, we shall focus on the modules $Q_{k,-q^r}$ only.
Let us then introduce an alternative parameterisation for the modules~\eqref{eq:Q=W/W}. For any positive integers $\lambda$ and $\lambdab$, we define
\begin{equation}
  [\lambda,\lambdab] := Q_{(\lambda-\lambdab)/2,-q^{(\lambda+\lambdab)/2}} \,.
\end{equation}
Note that, in particular, the vacuum module is given by\footnote{Note that, when $q$ is a root of unity, one still has $V=W_{0,-q}/W_{1,-1}$, but $V$ is no longer irreducible.} $V=Q_{0,-q}=[1,1]$. Moreover, one has the basic properties
\begin{equation} \label{eq:f.and.th.on.Q}
  F = f_\lambda\,\id \,,
  \qquad \Fb = f_{\lambdab}\,\id \,,
  \qquad \theta = \theta_{\lambda,\lambdab}\,\id
  \qquad \text{on } [\lambda,\lambdab]
\end{equation}
where $f_\lambda=-q^\lambda-q^{-\lambda}$ and $\theta_{\lambda,\lambdab}=(-q^{1/2})^{\lambda^2-\lambdab^2}$. The reflection automorphism acts as $\rho([\lambda,\lambdab])=[\lambdab,\lambda]$.
\medskip

In the following, we study the structure of the subcategory $\cal Q$ generated by all the modules of the form $[\lambda,\lambdab]$ with positive integers $\lambda$ and $\lambdab$.

\subsection{Decomposition of the tensor product}
\label{sec:decomp}

The main goal of this section is to derive the decomposition formula
\begin{equation} \label{eq:decomp.QxQ}
  [\lambda,\lambdab]\otimes [\mu,\mub]
  \equiv \bigoplus_{\smallarray{\nu=|\lambda-\mu|+1}{\lambda+\mu+\nu\text{ odd}}}^{\lambda+\mu-1}
  \bigoplus_{\smallarray{\nub=|\lambdab-\mub|+1}{\lambdab+\mub+\nub\text{ odd}}}^{\lambdab+\mub-1} [\nu,\nub]
\end{equation}
for any pair of simple modules $[\lambda,\lambdab], [\mu,\mub] \in \cal Q$.
\medskip

For this, using a technical assumption on the basis states, we shall derive the relation \eqref{eq:decomp.QxQ} in the cases involving the ``fundamental'' modules $[2,1]$ and $[1,2]$:
\begin{subequations}\label{eq:decomp.fund}
  \begin{alignat}{2}
    & [\lambda,\lambdab]\otimes [2,1] \equiv [\lambda+1,\lambdab] \oplus [\lambda-1,\lambdab] \label{eq:decomp.21} &\qquad & \text{if } \lambda>1 \\
    & [\lambda,\lambdab]\otimes [1,2] \equiv [\lambda,\lambdab+1] \oplus [\lambda,\lambdab-1] \label{eq:decomp.12} && \text{if } \lambdab>1 \\
    & [1,2]\otimes [2,1] \equiv [2,2] \,. \label{eq:decomp.22}
  \end{alignat}
\end{subequations}
The relations \eqref{eq:decomp.12} and \eqref{eq:decomp.21} are related by $q\leftrightarrow 1/q$. Hence, we first give the detailed derivation of \eqref{eq:decomp.21}, and then we treat the special case \eqref{eq:decomp.22}. The general case of~\eqref{eq:decomp.QxQ} can be deduced straightforwardly  from \eqref{eq:decomp.fund} by recursion on $\mu$ and $\mub$.

\paragraph{The case of $[\lambda,\lambdab]\otimes [2,1]$.}
Recall that $[\lambda,\lambdab]=Q_{k,-q^r}$ and $[2,1]=Q_{1/2,-q^{3/2}}$. In this section, we use the short-hand notation for the LHS of \eqref{eq:decomp.21}
\begin{equation}
  X := [\lambda,\lambdab]\otimes [2,1] = Q_{k,-q^r} \otimes Q_{1/2,-q^{3/2}} \,.
\end{equation}
Let us first argue that the dimensions of vector spaces for the LHS and RHS of~\eqref{eq:decomp.21} coincide.
For any link state of $X(N)$, denote respectively by $a$ and $b$ the marked points originating from $[\lambda,\lambdab]$ and $[2,1]$.
Due to the null-state conditions in $[\lambda,\lambdab]$ and $[2,1]$, the vector space $X(N)$ is spanned by link states with strictly less than $(|r|-k)$ loop segments crossing the reference line connected to $a$, and no arch separating the points $a$ and $b$. We assume that all such link states are linearly independent, and thus that they form a basis of $X(N)$.
\medskip

If $k>0$, the marked points $a$ and $b$ can be connected by zero or one leg. Due to the constraint on crossings with the reference line, the corresponding numbers of link states are
\begin{equation}
  d_{k+1/2}(N)-d_{|r|+1/2}(N)
  \qquad \text{and} \qquad
  d_{k-1/2}(N)-d_{|r|-1/2}(N) \,.
\end{equation}
If $k=0$, the leg attached to $b$ can have zero or one crossing with the reference line connected to $a$, and the corresponding numbers of link states are
\begin{equation}
  d_{1/2}(N)-d_{|r|+1/2}(N)
  \qquad \text{and} \qquad
  d_{1/2}(N)-d_{|r|-1/2}(N) \,.
\end{equation}
Hence, in both cases, we obtain
\begin{alignat}{1} 
  \dim X(N) &=
  [d_{k+1/2}(N)-d_{r+1/2}(N)]
  + [d_{k-1/2}(N)-d_{r-1/2}(N)] \nn\\
  &= \dim Q_{k+1/2,-q^{r+1/2}}(N) +  \dim Q_{k-1/2,-q^{r-1/2}}(N) \,.
  \label{eq:dim.X}
\end{alignat}
\medskip

We then introduce the states $w_m \in X(2m)$
\begin{equation} \label{eq:w_m}
  w_m = \begin{cases}
    c_0^{\dag m-k-1/2}u_k \otimes u_{1/2} & \text{if } m\geq k+\frac12 \\
    c_0^{k+1/2-m}\cdot (u_k\otimes u_{1/2}) & \text{if } m\leq k+\frac12
  \end{cases}
\end{equation}
for any non-negative $m\in \Zbb+k+1/2$. These states satisfy $w_m \propto c_0 \cdot w_{m+1}$,
and hence they generate nested submodules, i.e.
\begin{equation}
  X_m(N) := \TL^a(N,2m) \cdot w_m \,,
  \qquad X_m(N)\subseteq X_{m+1}(N) \,.
\end{equation}
Moreover, the space $X(N)$ is generated by the action of $\TL^a(N)$ on $w_{N/2}$ (see \cite{IMD24}), and hence $X_{N/2}(N)=X(N)$.
On the other hand, the null-state condition in the module $[2,1]$ reads
\begin{equation} \label{eq:null}
  (\cd_0-q^{1/2} \cd_1 - q^{-1/2}\cd_2)\cdot u_{1/2} = 0 \,.
\end{equation}
Consider the state $w_m$ with $m\geq k+3/2$. Using the identity $\cd_0 u\otimes v=\Omega^{-1}\cdot (u\otimes \cd_0 v)$ in any tensor product $A\otimes B$, we can write
\begin{alignat}{1}
  w_m &= 
   \Omega^{-1}\cdot (c_0^{\dag m-k-3/2}u_k\otimes \cd_0 u_{1/2})
  = \Omega^{-1}\cdot \left(c_0^{\dag m-k-3/2}u_k\otimes (q^{1/2} \cd_1 + q^{-1/2}\cd_2)u_{1/2}\right) \nn\\
  &= \Omega^{-1} (q^{1/2} \cd_{2m-2} + q^{-1/2}\cd_{2m-1}) \cdot w_{m-1} \,,
\end{alignat}
and hence $X_m=X_{m-1}$ for any $m\geq k+3/2$.  If $k\geq 3/2$, with a similar argument, one can show that $X_{k-3/2}=0$.
We thus have the filtration
\begin{equation} \label{eq:filtration}
  0=X_{k-3/2}\subseteq X_{k-1/2}\subseteq X_{k+1/2} = X \,,
\end{equation}
where, by convention, $X_m=0$ if $m$ is negative.
\medskip

The null-state condition \eqref{eq:null} yields the identity in $X$:
\begin{equation}
  u_k \otimes (\cd_0-q^{1/2} \cd_1 - q^{-1/2}\cd_2)\cdot u_{1/2} = 0 \,.
\end{equation}
Let us first discuss the case when $k\geq 1$.
Applying $c_0$ and $c_0^2$ on the above identity, we obtain respectively
\begin{equation} \label{eq:seed.X}
  (\Omega +q^{r+1/2}\id) \cdot w_{k+1/2}\propto \cd_{2k}\cdot w_{k-1/2} \,,
  \qquad
  (\Omega +q^{r-1/2}\id) \cdot w_{k-1/2}=0 \,.
\end{equation}
Moreover, for any $m\geq 1$ we have $c_0\cdot w_m=0$ in $X_m/X_{m-1}$. This proves that $w_{k+1/2}$ (resp. $w_{k-1/2}$) is a seed state of type $(k+1/2,-q^{r+1/2})$ [resp. $(k-1/2,-q^{r-1/2})$] in $X_{k+1/2}/X_{k-1/2}$ (resp. $X_{k-1/2}$).
\medskip

For $k=1/2$, instead of \eqref{eq:seed.X}, we obtain
\begin{equation}
  (\Omega +q^{r+1/2}\id) \cdot w_1 \propto \cd_1\cdot w_0 \,,
  \qquad
  (f-f_{r-1/2}\id) \cdot w_0 =0 \,,
\end{equation}
where $f_\lambda$ is defined in \eqref{eq:f.and.th.on.Q}.
Hence $w_1,w_0$ are seed states of types $(1,-q^{r+1/2})$ and $(0,-q^{r-1/2})$ respectively.
\medskip

For $k=0$, we get a single relation
\begin{equation}
  (\Omega +q^{1/2+r}\id)(\Omega +q^{1/2-r}\id) \cdot w_{1/2} =0 \,.
\end{equation}
We then introduce the states
\begin{equation}
  w_{1/2}^+=(\Omega +q^{1/2-r}\id) \cdot w_{1/2} \,,
  \qquad w_{1/2}^-=(\Omega +q^{1/2+r}\id) \cdot w_{1/2} \,,
\end{equation}
so that each state $w_{1/2}^\pm$ is a seed state of type $(1/2,1/2\pm r)$ in $X_{1/2}$.
\medskip

Thus, in the three cases, we have constructed seed states of types $(k+1/2,r+1/2)$ and $(k-1/2,r-1/2)$ in the quotients of the filtration \eqref{eq:filtration}.
Using the identity of dimensions \eqref{eq:dim.X} and the fact that the corresponding values $f_{k+r+1}$ and $f_{k+r-1}$ \eqref{eq:f.and.th.on.Q} for the central element $F$ are distinct, this proves the decomposition \eqref{eq:decomp.21}.

\paragraph{The case of $[1,2]\otimes [2,1]$.}
We can keep the notations introduced for $[\lambda,\lambdab]\otimes[2,1]$, and set $k=1/2$ and $r=-3/2$.
For clarity, we denote by $u_{1/2}$ and $\ub_{1/2}$ the unique link states of $Q_{1/2,-q^{3/2}}(1)$ and $Q_{1/2,-q^{-3/2}}(1)$, respectively.
Using the identities for any $u\in A(L)$ and $v\in B(M)$
\begin{equation}
  \Omega \cdot (u\otimes v) = c_L\cdot (\Omega u\otimes \cd_0 v) \,,
  \qquad
  \Omega^{-1} \cdot (u\otimes v) = c_{L+2}\cdot (\cd_0 u\otimes  \Omega^{-1} v) \,,
\end{equation}
we obtain
\begin{equation}
  \begin{aligned}
    & \Omega\, w_1 = -q^{-3/2} c_1\cdot (\ub_{1/2}\otimes \cd_0 u_{1/2})
    = -q^{-1} \,w_1- q\, \cd_1\,w_0 \\
    & \Omega^{-1}\, w_1 = -q^{3/2} c_3\cdot (\cd_0 \ub_{1/2}\otimes u_{1/2})
    = -q^2\, w_1 -q^4\, \cd_1\,w_0 \,,
  \end{aligned}
\end{equation}
where we have used the null-state conditions on $u_{1/2}$ and $\ub_{1/2}$, respectively.
Combining the above relations, we get
\begin{equation}
  w_1 = \frac{q^2\cd_0-q \cd_1}{q^{-1}-q^{-2}} \cdot w_0 \,,
\end{equation}
and hence $X_1=X_0$.
Moreover, like in the previous case with $k=1/2$, we have $f\cdot w_0 = (-q^2-q^{-2})\,w_0$, and hence $w_0$ is a seed state of type $(0,-q^2)$ in $X$.
Hence, $X$ is spanned by the link states where $a$ and $b$ are connected by a leg, and with at most one loop segment crossing the reference line connected to $a$. We assume that these states are linearly independent. An argument similar to the previous case then leads to the proof of \eqref{eq:decomp.22}.

\paragraph{The general case.}
The decomposition \eqref{eq:decomp.QxQ} for general $\mu,\mub$ is obtained by a recursive argument based on \eqref{eq:decomp.fund}, and using the associativity of the tensor product $\otimes$. Let us illustrate the argument on the example $[\lambda,\lambdab]\otimes[3,1]$, with $\lambda>2$. Using \eqref{eq:decomp.21}, we have
\begin{equation}
  [2,1]\otimes [2,1] \equiv [1,1] \oplus [3,1]
\end{equation}
and thus
\begin{equation}
  [\lambda,\lambdab]\otimes[2,1]\otimes [2,1]
  \equiv [\lambda,\lambdab] \oplus [\lambda,\lambdab]\otimes [3,1]
\end{equation}
where we have used the fact that $[1,1]=V$.
On the other hand, using \eqref{eq:decomp.21} twice, we obtain
\begin{equation}
  [\lambda,\lambdab]\otimes[2,1]\otimes [2,1]
  \equiv [\lambda+2,\lambdab] \oplus 2[\lambda,\lambdab] \oplus [\lambda-2,\lambdab] \,.
\end{equation}
Comparing the two above relations, we obtain
\begin{equation}
  [\lambda,\lambdab]\otimes [3,1]
  \equiv [\lambda+2,\lambdab] \oplus [\lambda,\lambdab] \oplus [\lambda-2,\lambdab] \,,
\end{equation}
which corresponds to \eqref{eq:decomp.QxQ} for $[\mu,\mub]=[3,1]$.

\paragraph{Fusion numbers.} For any simple modules $[\lambda,\lambdab],[\mu,\mub]\in\cal Q$, the decomposition \eqref{eq:decomp.QxQ} can be written
\begin{equation}
  [\lambda,\lambdab]\otimes [\mu,\mub]
  \equiv \bigoplus_{\nu,\nub} n_{\lambda \mu}^\nu \, n_{\lambdab \mub}^{\nub} \, [\nu,\nub] \,,
\end{equation}
where
\begin{equation} \label{eq:n}
  n_{\lambda \mu}^\nu := \begin{cases}
    1 & \text{if } \nu \in \{|\lambda-\mu|+1, |\lambda-\mu|+3,\dots, \lambda+\mu-1 \} \\
    0 & \text{otherwise.}
  \end{cases}
\end{equation}
For any triplet $A=[\lambda,\lambdab],B=[\mu,\mub],C=[\nu,\nub]\in\cal Q$, the multiplicity of $C$ in $A\otimes B$, called the fusion number $N_{AB}^C$, is given by
\begin{equation}
  N_{AB}^C = N_{[\lambda,\lambdab],[\mu,\mub]}^{[\nu,\nub]}
  = n_{\lambda \mu}^\nu \, n_{\lambdab \mub}^{\nub} \,.
\end{equation}

\paragraph{Seed states.}
Consider the tensor product $A\otimes B$, where $A=Q_{k,-q^r}=[\lambda,\lambdab]$ and $B=Q_{\ell,-q^s}=[\mu,\mub]$, with $k,\ell\in\Zbb_{\geq 0}/2$.
Generalising from the case $[\mu,\mub]=[2,1]$, let us introduce the states
\begin{equation} \label{eq:w.klm}
  w_{k\ell}^m = \begin{cases}
    c_0^{\dag m-k-\ell}u_k \otimes u_\ell & \text{if } m\geq k+\ell \\
    c_0^{k+\ell-m}\cdot (u_k\otimes u_\ell) & \text{if } m\leq k+\ell
  \end{cases}
\end{equation}
and the submodules $X_m(N)=\TL^a(N,2m)\cdot w_{k\ell}^m$, 
for any non-negative $m$ such that $m+k+\ell\in \Zbb$.
We then have $w_{k,\ell}^m\propto c_0\cdot w_{k,\ell}^{m+1}$, and $(A\otimes B)(N)=\TL^a(N)\cdot w_{N/2}$, which yields
\begin{equation}
  X_m(N)\subseteq X_{m+1}(N) \subseteq \dots \subseteq X_{N/2}(N)=(A\otimes B)(N) \,.
\end{equation}
Let $C=[\nu,\nub]=Q_{m,-q^t}$ such that $N_{AB}^C=1$. Using \eqref{eq:decomp.QxQ}, the seed state of type $(m,-q^t)$ in $A\otimes B$ can be written
\begin{equation}
  \xi_{AB}^C = \pi_{AB}^C \cdot w_{k\ell}^m \,,
\end{equation}
where $\pi_{AB}^C$ is the projector on the eigenspace $(F,\Fb)=(f_\nu,f_{\nub})$ in $X_m$. This projector can be expressed as a divisor of the polynomial
\begin{equation}
  \prod_{\smallarray{\alpha=|\lambda-\mu|+1}{\lambda+\mu+\alpha \text{ odd and } \alpha\neq \nu}}^{\lambda+\mu-1}
  \frac{(F-f_\alpha)}{(f_\nu-f_\alpha)}
  \prod_{\smallarray{\bar\alpha=|\lambdab-\mub|+1}{\lambdab+\mub+\bar\alpha \text{ odd and } \bar\alpha\neq \nub}}^{\lambdab+\mub-1}
  \frac{(\Fb-f_{\bar\alpha})}{(f_{\nub}-f_{\bar\alpha})} \,,
\end{equation}
where $f_\lambda$ is defined in \eqref{eq:f.and.th.on.Q}.
For instance, for $A=B$ we have
\begin{equation} \label{eq:decomp.AxA}
  [\lambda,\lambdab]\otimes [\lambda,\lambdab]
  \equiv \bigoplus_{i=1}^{\lambda}
  \bigoplus_{j=1}^{\lambdab} [2i-1,2j-1] \,,
\end{equation}
and we can write the seed state for $V=[1,1]$ in $A\otimes A$ as
\begin{equation} \label{eq:xi.AAV}
  \xi_{AA}^V = \pi_{AA}^V \cdot w_{kk}^0 \,,
  \qquad \pi_{AA}^V =
  \prod_{i=2}^{\mathrm{min}(\lambda,\lambdab)}
  \frac{(F-f_{2i-1})}{(\beta-f_{2i-1})} \,.
\end{equation}

\paragraph{Invariant bilinear form on $(A\otimes B)\times \Cb$.}
Consider three simple modules $A=Q_{k,-q^r}$, $B=Q_{\ell,-q^s}$ and $C=Q_{m,-q^t}$ in $\cal Q$, such that $N_{AB}^C=1$, and denote $\Cb=\rho(C)=Q_{m,-q^{-t}}$. By construction, the Gram form $\aver{\,,\,}$ on $\Cb\times C$ is non-degenerate. Hence, from~\eqref{eq:Hom=F}, the space $\I(A\otimes B, \Cb)$ of invariant bilinear forms on $(A\otimes B)\times \Cb$ is one-dimensional.
\medskip

Let $g$ be a non-zero element of $\I(A\otimes B, \Cb)$, and let us show that $g(w_{k\ell}^m,\ub_m)\neq 0$, where $\ub_m$ denotes the unique link state of $\Cb(2m)$.
The argument simply relies on the identities
\begin{equation}
  \Cb(N)=\TL^a(N,2m)\cdot \ub_m \,,
  \qquad \Cb(2m)=\mathrm{span}(\ub_m) \,,
  \qquad (A\otimes B)(2m)=\TL^a(2m)\cdot w_{k\ell}^m \,.
\end{equation}
Since $g\neq 0$, there exists an integer $N$ and two states $u\in (A\otimes B)(N)$ and $v\in \Cb(N)$ such that $g(u,v)\neq 0$. On the other hand, we can write $v=\lambda\cdot \ub_m$ with $\lambda\in\TL^a(N,2m)$, and thus
\begin{equation}
  g(u,v) = g(u,\lambda\cdot \ub_m) = g(\lambda^t\cdot u,\ub_m) \,.
\end{equation}
Since $\lambda^t\cdot u\in (A\otimes B)(2m)$, we write it as $\lambda^t\cdot u=\mu\cdot w_{k\ell}^m$ with $\mu\in\TL^a(2m)$, which yields
\begin{equation}
  g(u,v) = g(\mu\cdot w_{k\ell}^m,\ub_m)
  = g(w_{k\ell}^m,\mu^t\cdot \ub_m) \propto g(w_{k\ell}^m,\ub_m) \,,
\end{equation}
and hence $g(w_{k\ell}^m,\ub_m) \neq 0$.
\medskip

Therefore, we define $\aaver{\,,\,}$ as the unique invariant bilinear form on $(A\otimes B)\times \Cb$ such that $\aaver{w_{k\ell}^m,\ub_m} =1$.
With a similar argument, there exists a unique invariant bilinear form on $\Cb\times (A\otimes B)$, which we also denote $\aaver{\,,\,}$, such that $\aaver{\ub_m,w_{k\ell}^m} =1$.

\section{Duality, braiding and associativity matrices}
\label{sec:Q2}

In this section, we consider the category $\cal Q$ spanned by the simple quotients of reducible standard modules of $\TL^a$.
We argue that any simple module $A=[\lambda,\lambdab]\in \cal Q$ is its own left and right dual, by constructing the evaluation and coevaluation morphisms $\eps_A:A\otimes A\to V$ and $\eta_A:V\to A\otimes A$. This result, together with the simple properties of the braiding, shows that the subcategory $\cal Q$ has a ribbon structure.
Finally, we examine the associativity matrices $\F_{ABC}^D$, and we show that they are given by the tensor product of two $U_q(sl_2)$ associativity matrices. 
The results are valid only when $q$ is not a root of unity.

\subsection{Duality}

Let $A=Q_{k,-q^r}=[\lambda,\lambdab]\in \cal Q$.
In this section, we construct a pair of morphisms $\eta_A:V\to A\otimes A$ and $\eps_A:A\otimes A\to V$ such that
\begin{equation} \label{eq:duality.cond}
  (\id_A \otimes \eps_A) \cdot (\eta_A \otimes \id_A) = \id_A \,,
  \qquad
  (\eps_A \otimes \id_A) \cdot (\id_A \otimes \eta_A) = \id_A \,,
\end{equation}
and we calculate the quantum dimension $d_A$, defined as
\begin{equation}
  \eps_A \cdot \eta_A = d_A\, \id_V \,.
\end{equation}

\paragraph{The morphisms $\eta_A$ and $\eps_A$.} Since $V=Q_{0,-q}$, the morphism $\eta_A$ can be defined as the insertion map \eqref{eq:insertion} associated to the seed state $\xi_{AA}^V$ \eqref{eq:xi.AAV}. It is convenient to introduce the unnormalised morphism $\wh\eta_A$, defined as
\begin{equation}
  \wh\eta_A(\lambda\cdot v_0) := \lambda\cdot \xi_{AA}^V
\end{equation}
for any $\lambda\in \TL^a(N,0)$, where $v_0$ is the unique link state of $V(0)$.
\medskip

For $\eps_A$, we use \eqref{eq:Hom=F}. We take any pair of mutually dual bases $(\vtheta_i),(\vthetab_i)$ of $V(N)$, and we define the unnormalised morphism
\begin{equation}
  \wh\eps_A(u) := \sum_i \aaver{u,\vthetab_i}\,\vtheta_i 
\end{equation}
for any $u\in (A\otimes A)(N)$.
\medskip

The calculation of $\wh\eps_A\cdot \wh\eta_A$ is straightforward. Since $V(N)=\TL^a(N,0)\cdot v_0$, it is sufficent to compute $(\wh\eps_A\cdot \wh\eta_A)(v_0)$. We write
\begin{equation}
  (\wh\eps_A\cdot \wh\eta_A)(v_0) = \wh\eps_A(\xi_{AA}^V)
  = \pi_{AA}^V\cdot \wh\eps_A(w_{kk}^0) = \aaver{w_{kk}^0,v_0}\,,\pi_{AA}^V\cdot v_0 = v_0 \,,
\end{equation}
and thus we have $\wh\eps_A\cdot \wh\eta_A = \id_A$.

\paragraph{Overlaps.} The next step is to calculate $(\id_A \otimes \wh\eps_A) \cdot (\wh\eta_A \otimes \id_A)$ and $(\wh\eps_A \otimes \id_A) \cdot (\id_A\otimes \wh\eta_A )$.
Since $A=\TL^a(N,2k)\cdot u_k$, it is sufficient to compute the overlaps
\begin{equation} \label{eq:overlaps.eps.eta}
  \aaver{\ub_k,(\id_A \otimes \wh\eps_A) \cdot (\wh\eta_A \otimes \id_A)(u_k)} \,,
  \qquad \aaver{\ub_k,(\wh\eps_A \otimes \id_A) \cdot (\id_A\otimes \wh\eta_A )(u_k)}
\end{equation}
in $\Ab\times (A\otimes V)$ and $\Ab\times (V\otimes A)$, respectively.
Let us focus on the first of these quantities.
We write the state $w_{kk}^0$ as
\begin{equation}
  w_{kk}^0 = c_0^{2k}\cdot (u_k\otimes u_k) = c_1\dots c_{2k}\cdot (u_k\otimes u_k) \,,
\end{equation}
and the coefficients of the polynomial $\pi_{AA}^V$ \eqref{eq:xi.AAV} as
\begin{equation}
  \pi_{AA}^V = \sum_{n=0}^{\mathrm{min}(\lambda,\lambdab)-1} \pi_{A,n}\, f^n \,.
\end{equation}
We then have
\begin{alignat}{1}
  & \aaver{\ub_k,(\id_A \otimes \wh\eps_A) \cdot (\wh\eta_A \otimes \id_A)(u_k)}
  = \aaver{\ub_k,(\id_A \otimes \wh\eps_A) \cdot (\xi_{AA}^V \otimes u_k)} \nn\\
  &\qquad= \aaver{\ub_k,(\id_A \otimes \wh\eps_A) \cdot (\pi_{AA}^Vc_1\dots c_{2k}(u_k\otimes u_k) \otimes u_k)} \nn\\
  &\qquad= \sum_{n=0}^{\mathrm{min}(\lambda,\lambdab)-1} \pi_{A,n} \, \aaver{\ub_k,(\id_A \otimes \wh\eps_A) \cdot (f^n\,c_1\dots c_{2k}(u_k\otimes u_k) \otimes u_k)} \nn\\
  &\qquad= \sum_{n=0}^{\mathrm{min}(\lambda,\lambdab)-1} \pi_{A,n} \, \aaver{\ub_k,c_0^n\,c_1\dots c_{2k}(\id_A \otimes \wh\eps_A) \cdot (u_k\otimes u_k \otimes c_0^{\dag n} u_k)} \nn\\
  &\qquad= \sum_{n=0}^{\mathrm{min}(\lambda,\lambdab)-1} \pi_{A,n} \, \aaver{c_0^{\dag n}\,\ub_k,c_1\dots c_{2k}(u_k \otimes \wh\eps_A(u_k \otimes c_0^{\dag n} u_k))} \nn\\
  &\qquad= \sum_{n=0}^{\mathrm{min}(\lambda,\lambdab)-1}\sum_i \pi_{A,n} \, \aaver{c_0^{\dag n}\,\ub_k,c_1\dots c_{2k}(u_k \otimes \vtheta_i)} \aaver{u_k \otimes c_0^{\dag n} u_k,\vthetab_i} \nn\\
  &\qquad= \sum_{n=0}^{\mathrm{min}(\lambda,\lambdab)-1}\sum_i \pi_{A,n} \, \aaver{\ub_k\otimes c_0^{\dag n}\,\ub_k,\vtheta_i} \aaver{u_k \otimes c_0^{\dag n} u_k,\vthetab_i} \nn\\
  &\qquad= \sum_{n=0}^{\mathrm{min}(\lambda,\lambdab)-1}\sum_i \pi_{A,n} \, \aaver{\Omega^n \wb_{kk}^{2k+n},\vtheta_i} \,\aaver{\Omega^n w_{kk}^{2k+n},\vthetab_i} \label{eq:calc.overlap}
\end{alignat}
where $\wb_{k\ell}^m$ denotes the analog of $w_{k\ell}^m$ in $\Ab\otimes \Bb$.
Here, we have used the relation $fu\otimes v = c_0\cdot(u\otimes \cd_0 v)$, and
the identity
\begin{equation}
  \aaver{u,c_1\dots c_M(v\otimes w)} = \aaver{u\otimes \rho(v), w}
\end{equation}
for any $u\in \Ab(L)$, $v\in A(M)$ and $w\in V(L+M)$.
Since $\wh\eps_A$ does not depend on the choice of mutually dual bases $(\vtheta_i),(\vthetab_i)$, we can change $\vtheta_i\to\Omega^n\vtheta_i$ and $\vthetab_i\to\Omega^n\vthetab_i$. This yields
\begin{equation} \label{eq:overlap=sum}
  \aaver{\ub_k,(\id_A \otimes \wh\eps_A) \cdot (\wh\eta_A \otimes \id_A)(u_k)}
  = \sum_{n=0}^{\mathrm{min}(\lambda,\lambdab)-1}\sum_i \pi_{A,n} \, \aaver{\wb_{kk}^{2k+n},\vtheta_i} \, \aaver{w_{kk}^{2k+n},\vthetab_i}
\end{equation}
for any pair of mutually dual bases $(\vtheta_i),(\vthetab_i)$ of $V(4k+2n)$.
A similar calculation for the second overlap in \eqref{eq:overlaps.eps.eta} yields exactly the same result.

\paragraph{Quantum dimension and quantum trace.}
Deriving a general expression for the sum in \eqref{eq:overlap=sum} is a non-trivial exercise, involving the enumeration of the basis $(\vtheta_i)$ and the inversion of the Gram matrix. Instead, we use computer algebra to determine this sum, for values of $\lambda$ and $\lambdab$ up to $7$, and we obtain the formula
\begin{equation} \label{eq:Z_AAV}
  \aaver{\ub_k,(\id_A \otimes \wh\eps_A) \cdot (\wh\eta_A \otimes \id_A)(u_k)}
  = \frac{1}{[\lambda]\, [\lambdab]}
\end{equation}
where $[m]$ is the $q$-integer
\begin{equation}
  [m]=[m]_{-q} = \frac{(-q)^m - (-q)^{-m}}{-q+q^{-1}} \,.
\end{equation}
The crucial fact is that $(\id_A \otimes \wh\eps_A) \cdot (\wh\eta_A \otimes \id_A)=(\wh\eps_A \otimes \id_A) \cdot (\id_A \otimes \wh\eta_A) \neq 0$, and hence $\wh\eta_A$ and $\wh\eps_A$ can be rescaled to satisfy \eqref{eq:duality.cond}:
\begin{equation}
  \eta_A = \mathcal{C}_{AA}^V\, \wh\eta_A \,,
  \qquad \eps_A = \mathcal{C}^{AA}_V\, \wh\eps_A \,,
\end{equation}
with constants $\mathcal{C}_{AA}^V$ and $\mathcal{C}^{AA}_V$ subject to the condition
$\mathcal{C}_{AA}^V\, \mathcal{C}^{AA}_V = [\lambda]\, [\lambdab]$.
We thus obtain the expression for the quantum dimension:
\begin{equation}
  d_A = [\lambda]\, [\lambdab] \qquad \text{for } A=[\lambda,\lambdab] \,.
\end{equation}
The morphisms $\eta$ and $\eps$ easily extend to any $A \in \mathcal{Q}$, by $\eta_{A\oplus B}=\eta_A+\eta_B$, and $\eps_{A\oplus B}=\eps_A+\eps_B$. For any $A\in \mathcal{Q}$ and any morphism $\phi:A\to A$, one defines the quantum trace
\begin{equation}
  \mathrm{qtr}_A(\phi) := \aver{v_0,\eps_A \cdot (\phi \otimes \id_A) \cdot \eta_A(v_0)}
\end{equation}
where $v_0$ is the single basis state of $V(0)$.

\subsection{Splitting and fusing morphisms}

\paragraph{Definitions.}
Consider three simple modules in $\cal Q$, denoted
$$A = Q_{k,-q^r}=[\lambda,\lambdab] \,, \qquad B = Q_{\ell,-q^s}=[\mu,\mub] \,, \qquad C = Q_{m,-q^t}=[\nu,\nub] \,,$$
and such that $N_{AB}^C=1$.
Similarly to $\wh\eta_A$ and $\wh\eps_A$, we introduce the morphisms
\begin{equation}
  \wh\psi_{AB}^C: \begin{cases}
    C\to A\otimes B \\
    \lambda\,u_m\mapsto \lambda\, \xi_{AB}^C
  \end{cases}
  \qquad \wh\psi_C^{AB}: \begin{cases}A\otimes B \to C \\
    u\mapsto \sum_i \aaver{u,\yb_i}\, y_i
  \end{cases}
\end{equation}
for any $\lambda\in \TL^a(N,2m)$ and $u\in (A\otimes B)(N)$.
Here, $(y_i),(\yb_i)$ denote any pair of mutually dual bases of $C(N)$ and $\Cb(N)$.
Since $C$ is simple, we have $\wh\psi_C^{AB}\cdot \wh\psi_{AB}^C\propto \id_C$, and the corresponding overlap reads:
\begin{equation}
  \aver{\ub_m,\wh\psi_C^{AB}\cdot \wh\psi_{AB}^C(u_m)}
  = \aver{\ub_m,\wh\psi_C^{AB}(\xi_{AB}^C)}
  = \aver{\ub_m,\wh\psi_C^{AB}(w_{k\ell}^m)}
  = \aver{\ub_m,u_m}\, \aaver{w_{k\ell}^m,\ub_m}=1 \,,
\end{equation}
which yields $\wh\psi_C^{AB}\cdot \wh\psi_{AB}^C=\id_B$.
We define the rescaled morphisms
\begin{equation} \label{eq:def.psi}
  \psi_{AB}^C := \mathcal{C}_{AB}^C\, \wh\psi_{AB}^C \,,
  \qquad \psi^{AB}_C := \mathcal{C}^{AB}_C\, \wh\psi^{AB}_C \,,
\end{equation}
with the constants $\mathcal{C}_{AB}^C$ and $\mathcal{C}^{AB}_C$ subject to the condition
\begin{equation} \label{eq:CC=sqrt}
  \mathcal{C}_{AB}^C\,\mathcal{C}^{AB}_C = \sqrt{d_Ad_B/d_C} \,.
\end{equation}
These morphisms then satisfy the relations 
\begin{equation} \label{eq:psi.rel}
  \psi_C^{AB}\cdot \psi_{AB}^C= \sqrt{\frac{d_Ad_B}{d_C}}\, \id_C \,,
  \qquad \id_{A\otimes B} = \sum_{N_{AB}^C=1} \sqrt{\frac{d_C}{d_Ad_B}}\, \psi^C_{AB}\cdot \psi^{AB}_C \,.
\end{equation}

\paragraph{Choice of gauge.}
Consider the operator which ``rotates'' a morphism:
\begin{equation}
  U:\left\{\begin{array}{ccc}
    \mathrm{Hom}(A\otimes B, C) &\to& \mathrm{Hom}(B\otimes C,A) \\
    \psi &\mapsto& (\id_A\otimes \eps_C) \cdot (\id_A\otimes \psi\otimes \id_C)  \cdot (\eta_A\otimes \id_B\otimes \id_C) \,.
  \end{array}\right.
\end{equation}
Recall that $\wh\psi^{AB}_C$ is related to the bilinear form $\aaver{\,,\,}\in\I(A\otimes B,\Cb)$ through the isomorphism \eqref{eq:Hom=F}. On the other hand, by combining \eqref{eq:Kr} and \eqref{eq:Kl}, we get the isomorphism of vector spaces
\begin{equation}
  \left\{\begin{array}{ccc}
    \I(A\otimes B,\Cb) &\to& \I(B\otimes C,\Ab) \\
    g &\mapsto& \check{g}
  \end{array}\right.
\end{equation}
where
\begin{equation}
  \check{g}(v\otimes w,\ub)= g\left(c_{N+1}\dots c_{N+M}(\rho(\ub)\otimes v),\rho(w)\right)
\end{equation}
for any $\ub\in \Ab(M+N)$, $v\in B(M)$ and $w\in C(N)$. Hence, there exists a non-zero constant $\omega_{ABC}$ such that
\begin{equation}
  \aaver{c_{N+1}\dots c_{N+M}(\rho(\ub)\otimes v),\rho(w)} = \omega_{ABC} \aaver{v\otimes w,\ub}
\end{equation}
for any $\ub,v,w$ as defined above, and the explicit expression of the constant $\omega_{ABC}$ can be obtained through a simple computation involving the states $w_{k\ell}^m$ (see App.~\ref{app:form}).
A natural choice of gauge for the morphisms \eqref{eq:def.psi} is then to impose the condition
\begin{equation} \label{eq:R.psi}
  U(\psi^{AB}_C) = \omega_{ABC} \, \psi^{BC}_A \,,
\end{equation}
i.e. that under a rotation, the morphisms pick the same factors as the bilinear forms.
\medskip

It is convenient to decompose the rotation as $U=U_r^{-1}\cdot U_\ell$, where
\begin{equation}
  U_\ell(\psi)=(\id_A\otimes \psi)\cdot (\eta_A\otimes \id_B) \,,
  \qquad U_r(\psi)=(\psi\otimes \id_B)\cdot(\id_A\otimes \eta_B)
\end{equation}
for any $\psi \in \mathrm{Hom}(A\otimes B,C)$.
The corresponding relations on bilinear forms read
\begin{equation}
  \begin{aligned}
    & \aaver{\vb,c_1\dots c_L\cdot(u\otimes w)}
    = a_{ABC}\, \aaver{\ub\otimes \vb,w} \,, \\
    & \aaver{\ub,c_{L+1}\dots c_{L+M} \cdot(w\otimes v)}
    = b_{ABC}\, \aaver{\ub\otimes \vb,w} \,,
  \end{aligned}
\end{equation}
for any $u\in A(L)$, $v\in B(M)$ and $w\in C(L+M)$.
Similarly to the case of $(\id_A\otimes \eps_A)\cdot(\eta_A\otimes \id_A)$, we
obtain the overlaps
\begin{equation}
  \begin{aligned}
    & \aver{\ub_\ell,\wh\psi_B^{AC}(\id_A\otimes \wh\psi^{AB}_C)(\wh\eta_A \otimes \id_B)(u_\ell)}
    = a_{ABC} \, Z_{ABC} \,, \\
    & \aver{\ub_k,\wh\psi_B^{AC}(\wh\psi^{AB}_C \otimes \id_B) (\id_A\otimes \wh\eta_B)(u_k)}
    = b_{ABC} \, \bar{Z}_{ABC} \,,
  \end{aligned}
\end{equation}
where
\begin{equation} \label{eq:Z_ABC}
  \begin{aligned}
    & Z_{ABC}= \sum_{n=0}^{\mathrm{min}(\lambda,\lambdab)-1} \sum_i (\pi_{AA}^V)_n\, \aaver{\wb_{k,\ell}^{k+\ell+n},y_i}  \, \aaver{w_{k,\ell}^{k+\ell+n}, \yb_i} \,, \\
    & \bar{Z}_{ABC} = \sum_{n=0}^{\mathrm{min}(\mu,\mub)-1} \sum_i (\pi_{BB}^V)_n\, \aaver{\wb_{k,\ell}^{k+\ell+n},y_i}  \, \aaver{w_{k,\ell}^{k+\ell+n}, \yb_i} \,,
  \end{aligned}
\end{equation}
and $(y_i),(\yb_i)$ is any pair of mutually dual bases of $C(N),\bar{C}(N)$ with $N=2k+2\ell+2n$.
If $\mathrm{min}(\mu,\mub)\leq \mathrm{min}(\lambda,\lambdab)$ then $\pi_{BB}^V$ is a divisor of $\pi_{AA}^V$, and thus we have $\xi_{BB}^V=\pi_{AA}\cdot w_{\ell\ell}^0$. In this case, one can replace $\pi_{BB}^V$ by $\pi_{AA}^V$ in the derivation of $\bar{Z}_{ABC}$, and one gets $\bar{Z}_{ABC}=Z_{ABC}$. An analogous argument holds in the case $\mathrm{min}(\mu,\mub)\geq \mathrm{min}(\lambda,\lambdab)$, and hence we have
\begin{equation}
  \bar{Z}_{ABC}=Z_{ABC}
\end{equation}
for any simple modules $A,B,C \in \cal Q$.
We use computer algebra to calculate the sum in \eqref{eq:Z_ABC} for several choices of $A,B,C$.
We observe that the quantity $Z_{ABC}/d_C$ is invariant under any permutation of $A,B,C$:
\begin{equation} \label{eq:X_ABC}
  \Omega_{ABC} = \frac{Z_{ABC}}{d_C} \,,
  \qquad \Omega_{ABC}=\Omega_{p(A),p(B),p(C)} \,,
  \qquad p\in\mathfrak{S}_3
\end{equation}
for any simple modules $A,B,C \in \cal Q$.
\medskip

Imposing the conditions
\begin{equation} \label{eq:lower}
  U_\ell(\psi^{AB}_C) = a_{ABC}\, \psi_{AC}^B \,,
  \qquad U_r(\psi^{AB}_C) = b_{ABC}\, \psi_{CB}^A 
\end{equation}
yields the relation \eqref{eq:R.psi}. 
The constants must then satisfy
\begin{equation}
  Z_{ABC}\, \mathcal{C}_{AA}^V\, \mathcal{C}^{AB}_C = \mathcal{C}_{AC}^B \,,
  \qquad Z_{ABC}\, \mathcal{C}_{BB}^V\, \mathcal{C}^{AB}_C = \mathcal{C}_{CB}^A \,.
\end{equation}
Using the condition \eqref{eq:CC=sqrt}, the symmetry \eqref{eq:X_ABC}, and the value $Z_{AAV}=1/d_A$ from \eqref{eq:Z_AAV}, we obtain a solution in the form
\begin{equation} \label{eq:C}
  \mathcal{C}^{AB}_C = (d_C\, \Omega_{ABC})^{-1/2} \,,
  \qquad \mathcal{C}_{AB}^C = \left(d_Ad_B\, \Omega_{ABC}\right)^{1/2} \,.
\end{equation}
This solution is not unique, since a change of gauge
\begin{equation} \label{eq:gauge}
  \mathcal{C}^{AB}_C \to \gamma_{ABC}\, \mathcal{C}^{AB}_C \,,
  \qquad \mathcal{C}_{AB}^C \to \gamma_{ABC}^{-1}\,\mathcal{C}_{AB}^C
\end{equation}
preserves the relations \eqref{eq:CC=sqrt} and \eqref{eq:R.psi}, if $\gamma_{ABC}\in \Cbb^\times$ is of the form $\gamma_{ABC}=\wh\gamma_{ABC}\kappa_A\kappa_B/\kappa_C$, with $\wh\gamma_{BAC}=\wh\gamma_{ACB}=\wh\gamma_{ABC}^{-1}$ for any $A,B,C$. If the latter condition is relaxed, then only \eqref{eq:CC=sqrt} is guaranteed in general.

\subsection{Braiding and topological $S$-matrix}

Throughout this section, we consider three simple modules $A,B,C\in\cal Q$, denoted
\begin{equation} \label{eq:ABC}
  A = Q_{k,x}=[\lambda,\lambdab] \,, \qquad B = Q_{\ell,y}=[\mu,\mub] \,, \qquad C = Q_{m,z}=[\nu,\nub] \,.
\end{equation}

\paragraph{Braiding.}
Let $A,B,C$ be as in \eqref{eq:ABC}, and such that $N_{AB}^C=1$. 
The action of $\sigma_{AB}$ on $\wh\psi_{AB}^C$ reads
\begin{equation} \label{eq:sigma.psi}
  \sigma_{AB} \cdot \wh\psi_{AB}^C = \sigma_{AB}^C \, \wh\psi_{BA}^C \,,
\end{equation}
where $\sigma_{AB}^C$ is a constant factor.
In the module $B\otimes A$, the seed state for $C$ is given by $\xi_{BA}^C=\pi_{BA}^C\cdot w_{\ell,k}^m$, and $\pi_{BA}=\pi_{AB}$ as central elements of $\TL^a$.
By computing explicitly $\sigma_{AB}(w_{k\ell}^m)$ in terms of $w_{\ell k}^m$, we obtain
\begin{equation}
  \sigma_{AB}^C=\begin{cases}
  y^{-2\ell}\, z^{m-k+\ell} 
  &\text{if } m\geq k+\ell \\
  (x/y)^{k+\ell-m}\, y^{-2\ell}\, z^{m-k+\ell}
  &\text{if } |k-\ell|\leq m\leq k+\ell \\
  (x/y)^{2\ell}\, y^{-2\ell}
  &\text{if } 0\leq m\leq k-\ell \\
  (x/y)^{2k}\, y^{-2\ell}\, z^{2m}
  &\text{if } 0\leq m\leq \ell-k \,.
  \end{cases}
\end{equation}
For the monodromy, we then get
\begin{equation}
  \sigma_{BA} \cdot \sigma_{AB} \cdot \psi_{AB}^C = x^{-2k} \, y^{-2\ell}\,z^{2m} \, \psi_{BA}^C \,,
\end{equation}
which is consistent with \eqref{eq:g.theta} and \eqref{eq:f.and.th.on.Q}.
In the particular case when $A=B=I_{k,x}$ and $C=I_{m,z}$, the expression of $\sigma_{AA}^C$ reduces to
\begin{equation} \label{eq:sigma.AAC}
  \sigma_{AA}^C = x^{-2k}\, z^m \,.
\end{equation}

\paragraph{Ribbon structure.}
The twist automorphism defined by $\theta_A=\Omega^N$ on $A(N)$ satisfies the following properties.
\begin{itemize}
\item For any $A,B\in\cal Q$:
  \begin{equation}
    \sigma_{BA}\cdot\sigma_{AB} = \theta_{A\otimes B} \cdot (\theta_A^{-1} \otimes \theta_B^{-1}) \,.
  \end{equation}
\item For any $A\in \cal Q$:
  \begin{equation}
    (\id_A \otimes \eps_A) \cdot \theta_A \cdot (\eta_A \otimes \id_A) =
    (\eps_A \otimes \id_A) \cdot \theta_A \cdot (\id_A \otimes \eps_A) = 
    \theta_A \,.
  \end{equation}
  \item For the vacuum module $V$, one has $\theta_V = \id_V$.
\end{itemize}
Therefore, $\cal Q$ possesses a ribbon structure.

\paragraph{Topological spin.}
Let $A=Q_{k,x}=[\lambda,\lambdab]$ be a simple module in $\cal Q$. The topological spin of $A$ is defined by
\begin{equation}
  \vtheta_A = (\id_A \otimes \eps_A) \circ (\sigma_{AA} \otimes \id_A) \circ (\id_A \otimes \eta_A) \,.
\end{equation}
Using \eqref{eq:sigma.psi}, \eqref{eq:psi.rel} and \eqref{eq:lower}, we  write
\begin{alignat}{1}
  \vtheta_A &= \sum_C N_{AA}^C \, \frac{\sqrt{d_C}}{d_A} \,  (\id_A \otimes \eps_A)  ((\sigma_{AA} \cdot \psi_{AA}^C \cdot \psi_C^{AA})\otimes \id_A) (\id_A \otimes \eta_A)  \nn \\
  &= \sum_C N_{AA}^C \, \frac{\sqrt{d_C}}{d_A} \, \sigma_{AA}^C
  (\id_A \otimes \eps_A)(\psi_{AA}^C\otimes \id_A)  (\psi_C^{AA}\otimes \id_A) (\id_A \otimes \eta_A) \nn \\
  &= \sum_C N_{AA}^C \, \frac{\sqrt{d_C}}{d_A} \, \sigma_{AA}^C \, \frac{b_{AAC}}{b_{CAA}}
  \,\psi_A^{CA} \cdot \psi_{CA}^A 
  = \sum_C N_{AA}^C \, \frac{d_C}{d_A} \, \sigma_{AA}^C \, \,\id_A \,.
  \label{eq:theta=sum}
\end{alignat}
Here, we have used $b_{AAC}=b_{CAA}$ from the symmetry \eqref{eq:sym.beta}.
Using the fusion rule \eqref{eq:decomp.AxA} and the expression \eqref{eq:sigma.AAC}, we obtain
\begin{equation}
  \vtheta_A = \frac{x^{-2k}}{[\lambda][\lambdab]}
  \sum_{\smallarray{\nu=1}{\nu\text{ odd}}}^{2\lambda-1} [\nu] e^{\frac{i\pi\nu}{2}} q^{\frac{\nu^2}{4}}
  \, \sum_{\smallarray{\nub=1}{\nub\text{ odd}}}^{2\lambdab-1} [\nub] e^{\frac{-i\pi\nub}{2}} q^{-\frac{\nub^2}{4}}
  \, \id_A = \theta_{\lambda,\lambdab} \, \id_A \,,
\end{equation}
where we used the identity
\begin{equation}
  \sum_{\smallarray{\nu=1}{\nu\text{ odd}}}^{2\lambda-1} [\nu] e^{\frac{i\pi\nu}{2}} q^{\frac{\nu^2}{4}}
  = [\lambda] \, i q^{\lambda^2-3/4}  \,.
\end{equation}
Therefore, the topological spin is identical to the twist automorphism.

\paragraph{Topological $S$-matrix.}
Let $A,B$ be as in \eqref{eq:ABC}.
The $S$-matrix element $S_{AB}$ is defined as
\begin{equation}
  S_{AB} := \mathrm{qtr}_{A\otimes B}(\sigma_{BA} \cdot \sigma_{AB}) \,.
\end{equation}
Note that $S$ is actually an infinite matrix, because $\cal Q$ has infinitely many simple modules.
Using the relation \eqref{eq:g.theta} and an argument similar to the case of $\vtheta_A$, we obtain
\begin{alignat}{1}
  S_{AB} &= \sum_C N_{AB}^C \, d_C \, \theta_C^{\phantom{-1}}\theta_A^{-1}\theta_B^{-1} \nn \\
  &= x^{-2k} y^{-2\ell}
  \sum_{\smallarray{\nu=|\lambda-\mu|+1}{\lambda+\mu+\nu\text{ odd}}}^{\lambda+\mu-1} [\nu](-q^{1/2})^{\nu^2}
  \, \sum_{\smallarray{\nub=|\lambdab-\mub|+1}{\lambdab+\mub+\nub\text{ odd}}}^{\lambdab+\mub-1} [\nub] (-q^{1/2})^{-\nub^2}\,.
\end{alignat}
Using the identity
\begin{equation}
  \sum_{\smallarray{\nu=|\lambda-\mu|+1}{\lambda+\mu+\nu\text{ odd}}}^{\lambda+\mu-1} [\nu](-q^{1/2})^{\nu^2}
  = [\lambda\mu] \, q^{\frac{\lambda^2+\mu^2-1}{2}} \, (-1)^{\lambda\mu} \,,
\end{equation}
this yields
\begin{equation}
  S_{AB} = [\lambda\mu]\, [\lambdab\mub] \,.
\end{equation}
The rescaled $S$-matrix element $\lambda_B(A)=S_{AB}/d_B$ is a fusion character, i.e.
\begin{equation}
  \lambda_B(A)\lambda_B(A') = \sum_{A''} N_{AA'}^{A''} \, \lambda_B(A'') \,.
\end{equation}
In particular, $\lambda_V(A) = d_A$, and $\lambda_B(V)=1$.

\subsection{Associativity matrices}

\paragraph{General case.}
Let $A,B,C,D$ be four simple modules in $\cal Q$.
Consider the two bases
$$\Big\{(\id_A\otimes \psi_{BC}^X)\cdot \psi_{AX}^D \Big\}_{N_{BC}^X=N_{AX}^D=1}
  \quad \text{and} \quad \Big\{(\psi_{AB}^Y \otimes \id_C)\cdot \psi_{YC}^D \Big\}_{N_{AB}^Y=N_{YC}^D=1}$$
for the morphism space $\mathrm{Hom}(D,A\otimes B\otimes C)$.
The matrix for the change of basis (also called $F$-symbol) is the associativity matrix $\F_{ABC}^D$. It is defined by the relation
\begin{equation}
  (\id_A\otimes \psi_{BC}^X)\cdot \psi_{AX}^D = \sum_Y \left(\F_{ABC}^D \right)_{YX} \, (\psi_{AB}^Y \otimes \id_C)\cdot \psi_{YC}^D  \,.
\end{equation}
Pictorially, this reads
\begin{equation}
  \mathfig{
    \begin{picture}(0,0)%
      \includegraphics{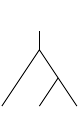}%
    \end{picture}
    \setlength{\unitlength}{3947sp}%
    \begin{picture}(630,1053)(4036,-2530)
      \put(4351,-1636){\makebox(0,0)[b]{\smash{\fontsize{11}{13.2}\normalfont {\color[rgb]{0,0,0}$D$}
      }}}
      \put(4426,-1936){\makebox(0,0)[lb]{\smash{\fontsize{11}{13.2}\normalfont {\color[rgb]{0,0,0}$X$}
      }}}
      \put(4051,-2461){\makebox(0,0)[b]{\smash{\fontsize{11}{13.2}\normalfont {\color[rgb]{0,0,0}$A$}
      }}}
      \put(4351,-2461){\makebox(0,0)[b]{\smash{\fontsize{11}{13.2}\normalfont {\color[rgb]{0,0,0}$B$}
      }}}
      \put(4651,-2461){\makebox(0,0)[b]{\smash{\fontsize{11}{13.2}\normalfont {\color[rgb]{0,0,0}$C$}
      }}}
    \end{picture}
  }
  \quad = \sum_Y \left(\F_{ABC}^D \right)_{YX} \quad
  \mathfig{
    \begin{picture}(0,0)%
      \includegraphics{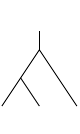}%
    \end{picture}
    \setlength{\unitlength}{3947sp}%
    \begin{picture}(630,1053)(4036,-2530)
      \put(4201,-2011){\makebox(0,0)[rb]{\smash{\fontsize{11}{13.2}\normalfont {\color[rgb]{0,0,0}$Y$}
      }}}
      \put(4351,-1636){\makebox(0,0)[b]{\smash{\fontsize{11}{13.2}\normalfont {\color[rgb]{0,0,0}$D$}
      }}}
      \put(4051,-2461){\makebox(0,0)[b]{\smash{\fontsize{11}{13.2}\normalfont {\color[rgb]{0,0,0}$A$}
      }}}
      \put(4351,-2461){\makebox(0,0)[b]{\smash{\fontsize{11}{13.2}\normalfont {\color[rgb]{0,0,0}$B$}
      }}}
      \put(4651,-2461){\makebox(0,0)[b]{\smash{\fontsize{11}{13.2}\normalfont {\color[rgb]{0,0,0}$C$}
      }}}
    \end{picture}
  }
\end{equation}
Let us denote:
$$A=Q_{k,-q^r} \,,\qquad C=Q_{\ell,-q^s} \,,\qquad D=Q_{m,-q^t} \,,\qquad X=Q_{\ell',-q^{s'}} \,.$$
Using \eqref{eq:psi.rel}, the associativity matrix element can be expressed as
\begin{equation}
  \left(\F_{ABC}^D \right)_{YX} = \sqrt{\frac{d_D}{d_Ad_Bd_C}}
  \, \aver{\ub_m,\psi_D^{YC}\cdot(\psi_Y^{AB}\otimes \id_C)\cdot(\id_A\otimes \psi_{BC}^X)\cdot\psi_{AX}^D(u_m)} \,.
\end{equation}
We fix the normalisation constants to \eqref{eq:C}.
This yields
\begin{equation} \label{eq:F=Z}
  \left(\F_{ABC}^D \right)_{YX} = \sqrt{\frac{d_X}{d_Y} \, \frac{\Omega_{BCX}\Omega_{AXD}}{\Omega_{YCD}\Omega_{ABY}}}
  \, \left(Z_{ABC}^D \right)_{YX} \,,
\end{equation}
where $\Omega_{ABC}$ is given by (\ref{eq:Z_ABC}--\ref{eq:X_ABC}), and $(Z_{ABC}^D)_{YX}$ is the overlap
\begin{equation}
  \left(Z_{ABC}^D \right)_{YX} = \aaver{\ub_m, (\wh\psi_Y^{AB}\otimes \id_C)\cdot(\id_A\otimes \wh\psi_{BC}^X)(w_{k\ell'}^m)} \,.
\end{equation}
Note that $\F_{ABC}^D$ is not gauge-invariant. Under the change of gauge \eqref{eq:gauge}, it is mapped into an equivalent matrix:
\begin{equation} \label{eq:gauge.F}
  \left(\F_{ABC}^D \right)_{YX} \to \frac{\gamma_{YCD} \,\gamma_{ABY}}{\gamma_{AXD} \,\gamma_{BCX}} \, \left(\F_{ABC}^D \right)_{YX} \,.
\end{equation}
\medskip

Before computing $\F_{ABC}^D$ explicitly, let us argue that it is sufficient to consider the cases $B=[2,1]$ and $B=[1,2]$ (actually, we shall focus on $B=[2,1]$, since $B=[1,2]$ is then easily obtained by $q\to 1/q$). Indeed, we can construct $\F_{ABC}^D$ recursively on $B$, using the fact that any simple module $B\in\cal Q$ can be obtained by iterating the fusion by $[2,1]$ and $[1,2]$. For instance, using the fusion rule
$$[2,1]\otimes [2,1]\equiv [1,1]\oplus[3,1]$$
one can obtain $\F_{A[3,1]C}^D$ from the matrices $\F_{A'[2,1]C'}^{D'}$ as follows.
Let us depict the module $[2,1]$ as a dot ($\bullet$), and $[3,1]$ as a square ($\scriptstyle\square$). The matrix $\F_{A[2,1]C}^D$ acts as
\begin{equation}
  \mathfig{\includegraphics{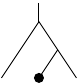}}
  \overset{\F}{\longrightarrow} \mathfig{\scalebox{-1}[1]{\includegraphics{F2.pdf}}}
\end{equation}
We then write
\begin{equation} \label{eq:rec.F}
  \mathfig{\includegraphics{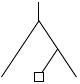}}
  \longrightarrow \mathfig{\includegraphics[scale=0.5]{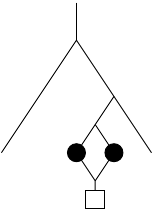}}
  \overset{\F^{-1}}{\longrightarrow} \mathfig{\includegraphics[scale=0.5]{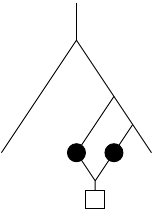}}
  \overset{\F}{\longrightarrow} \mathfig{\includegraphics[scale=0.5]{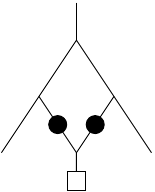}}
  \overset{\F}{\longrightarrow} \mathfig{\scalebox{-1}[1]{\includegraphics[scale=0.5]{FF3.pdf}}}
  \overset{\F^{-1}}{\longrightarrow} \mathfig{\scalebox{-1}[1]{\includegraphics[scale=0.5]{FF2.pdf}}}
  \longrightarrow \mathfig{\scalebox{-1}[1]{\includegraphics{FF1.pdf}}}
\end{equation}
where the first and last steps correspond to the relations \eqref{eq:psi.rel}.
\medskip

The results presented in the examples below suggest that, after an appropriate gauge transformation \eqref{eq:gauge.F}, the ``fundamental'' matrices $\F_{A[2,1]C}^D$ and $\F_{A[1,2]C}^D$ coincide with the $U_q(sl_2)$ associativity matrices (see Appendix~\ref{app:6j}):
\begin{equation}
  \F_{[\lambda,\lambda],[2,1],[\mu,\mub]}^{[\nu,\nub]} = F_{\lambda 2 \mu}^\nu \,,
  \qquad \F_{[\lambda,\lambdab],[1,2],[\mu,\mub]}^{[\nu,\nub]} = F_{\lambdab 2 \mub}^{\nub} \,.
\end{equation}
Hence, since the recursion relations like \eqref{eq:rec.F} only rely on the fusion rules and the fundamental $\F$-matrices, we obtain the factorised form
\begin{equation}
  \F_{[\lambda,\lambda],[\kappa,\bar\kappa],[\mu,\mub]}^{[\nu,\nub]}
  = F_{\lambda \kappa \mu}^\nu \otimes F_{\lambdab \bar\kappa \mub}^{\nub}  \,,
\end{equation}
where $\otimes$ stands for the usual tensor product of matrices.

\paragraph{The case of $B=[2,1]$.}
Suppose that
$$A=Q_{k,-q^r}=[\lambda,\lambdab] \qquad B=Q_{1/2,-q^{3/2}}=[2,1] \qquad C=Q_{\ell,-q^s}=[\mu,\mub] \qquad D=Q_{m,-q^t} \,,$$
with $0\leq k<|r|$, \ $1/2 \leq \ell <|s|$ and $0\leq m <|t|$.
We denote $A_\pm=[\lambda\pm 1,\lambdab]$ and $C_\pm=[\mu\pm 1,\mub]$, and we set $X=C_-=Q_{\ell-1/2,-q^{s-1/2}}$ and $Y\in\{A_+,A_-\}$.
Then we can write
\begin{equation}
  w_{k,\ell-1/2}^m = \begin{cases}
    c_0^{\dag(m-k-\ell+1/2)}u_k \otimes u_{\ell-1/2} & \text{if } m \geq k+\ell-1/2 \\
    c_0^{(k+\ell-m-1/2)}\cdot(u_k \otimes u_{\ell-1/2}) & \text{if } m \leq k+\ell-1/2 \\
  \end{cases}
\end{equation}
and
\begin{equation}
  \xi_{BC}^X = w_{1/2,\ell}^{\ell-1/2}=c_0 \cdot (u_{1/2} \otimes u_\ell) = q^{3/2-s} \, c_1 \cdot (u_{1/2} \otimes u_\ell) \,.
\end{equation}
\begin{itemize}
\item If $m \geq k+\ell-1/2$, we obtain
\begin{equation}
  \left(Z_{ABC}^D \right)_{YX} = q^{3/2-s} \sum_i \aaver{\ub_m,c_{2m-2\ell+2}\cdot(y_i\otimes u_\ell)}
  \, \aaver{c_0^{\dag(m-k-\ell+1/2)}u_k \otimes u_{1/2}, \yb_i} \,,
\end{equation}
where $(y_i),(\yb_i)$ is any pair of mutually dual bases of $Y(2m-2\ell+2),\Yb(2m-2\ell+2)$.
The first factor can be written
\begin{alignat}{1}
  \aaver{\ub_m,c_{2m-2\ell+2}\cdot(y_i\otimes u_\ell)}
  &= b_{DCY}\, \aaver{c_{2m-2\ell+2}\dots c_{2m}\cdot(\ub_m\otimes \ub_\ell),y_i} \nn \\
  &= b_{DCY}\, x\, \aaver{\wb_{m,\ell}^{m-\ell+1},y_i} \,,
\end{alignat}
where $x=q^{(t-s)(2\ell-1)}$. The second factor is equal to $\aaver{w_{k,1/2}^{m-\ell+1}, \yb_i}$.
\item If $m \leq k+\ell-1/2$, we write
\begin{equation}
  w_{k,\ell-1/2}^m
  = x \, c_{k-\ell+m+3/2}\dots c_{2k}\cdot (u_k\otimes u_{\ell-1/2}) \,,
\end{equation}
where $x=q^{(r-s+1/2)(k+\ell-m-1/2)}$,
which yields
\begin{alignat}{1}
  \left((Z_{ABC}^D \right)_{YX} &= q^{3/2-s} \,x\, \sum_i \aaver{\ub_m,c_{k-\ell+m+3/2}\dots c_{2k+1}\cdot(y_i\otimes u_\ell)}
  \, \aaver{u_k \otimes u_{1/2}, \yb_i} \nn \\
  &= q^{3/2-s} \,x\,x'\, b_{DCY}\, \sum_i \aaver{\wb_{m,\ell}^{k+1/2},y_i}
  \, \aaver{w_{k,1/2}^{k+1/2}, \yb_i}
\end{alignat}
where $x'=q^{(t-s)(\ell+m-k-1/2)}$ and
$(y_i),(\yb_i)$ is any pair of mutually dual bases of ${Y(2k+1)}$ and $\Yb(2k+1)$.
\end{itemize}

Hence, the overlap $(Z_{ABC}^D)_{YX}$ is given by
\begin{equation} \label{eq:Z_YX}
  (Z_{ABC}^D)_{YX} = b_{DCY} \, q^h\, \sum_i \aaver{\wb^n_{m,\ell},y_i} \, \aaver{w_{k,1/2}^n, \yb_i} \,,
\end{equation}
where
\begin{equation}
  \begin{aligned}
    & n = \begin{cases}
      m-\ell+1 & \text{if } m \geq k+\ell-1/2 \\
      k+1/2 & \text{if } m \leq k+\ell-1/2 \,, \\
    \end{cases} \\
    & h = (3/2-s)+(t-s)(2\ell-1)+(r-t+1/2)\mathrm{max}(k+\ell-m-1/2,0)
  \end{aligned}
\end{equation}
and $(y_i),(\yb_i)$ is any pair of mutually dual bases of $Y(2n),\Yb(2n)$.
In the following examples, we keep $B=[2,1]$, we fix $D$ to a given value, and we find the general form of $\F_{A[2,1]C}^D$ in terms of $A$ and $C$. For this, we use the expressions \eqref{eq:Z_ABC}, \eqref{eq:F=Z} and \eqref{eq:Z_YX}, and we resort to computer algebra to enumerate the bases $(y_i)$ and to determine the associated Gram matrices and their inverse.

\paragraph{Example 1: $D=[2,1]$.}
We set $A=[\lambda,\lambdab]$, $B=[2,1]$ and $D=[2,1]$. The space $\mathrm{Hom}(D,A\otimes B\otimes C)$ is two-dimensional if $C=A$, one-dimensional if $C=[\lambda\pm 2,\lambdab]$, and zero otherwise.
Let us study the case
$$A=C=Q_{k,-q^r}=[\lambda,\lambdab] \qquad B=Q_{1/2,-q^{3/2}}=[2,1] \qquad D=[2,1] \,.$$
with $1/2\leq k<|r|$.
We denote $A_\pm=[\lambda\pm 1,\lambdab]$, and we set $X=A_-$ and $Y\in\{A_+,A_-\}$.
Using computer algebra, we get the normalising factors (\ref{eq:Z_ABC}--\ref{eq:X_ABC})
\begin{equation}
  \begin{aligned}
    & \Omega_{[2,1][\lambda,\lambdab][\lambda+1,\lambdab]} = \frac{1}{[\lambda+1][\lambdab]}
    &\qquad& \Omega_{[2,1][\lambda,\lambdab][\lambda-1,\lambdab]} = \frac{1}{[\lambda][\lambdab]} \,,
  \end{aligned}
\end{equation}
and the overlaps \eqref{eq:Z_YX}
\begin{equation}
  \left(Z_{A[2,1]A}^{[2,1]} \right)_{A_+,A_-}
  = q^{1-r} \,,
  \qquad \left(Z_{A[2,1]A}^{[2,1]} \right)_{A_-,A_-}
  = \frac{(-1)^\lambda\, q^{1-\lambdab}}{[\lambda]} \,.
\end{equation}
This yields
\begin{equation}
  \left(\F_{A[2,1]A}^{[2,1]} \right)_{A_+,A_-}
  = q^{1-r} \sqrt{\frac{[\lambda+1][\lambda-1]}{[\lambda]^2}} \,,
  \qquad \left(\F_{A[2,1]A}^{[2,1]} \right)_{A_-,A_-}
  = \frac{(-1)^\lambda\, q^{1-\lambdab}}{[\lambda]} \,.
\end{equation}
The direct calculation of associativity matrix elements with $X=A_+$ is more technical, because $\xi_{BC}^X$ then involves a non-trivial projector $\pi_{BC}^X$ (see Sec.~\ref{sec:decomp}). Instead of this calculation, we use the fact that $\F_{A[2,1]A}^{[2,1]}$ must be unitary for $|q|=1$. In this case, the $q$-numbers are real, and we obtain a matrix of the form 
\begin{equation}
  \F_{A[2,1]A}^{[2,1]} = \left(\begin{array}{cc}
  \frac{(-1)^{\lambda+1}q^{1-\lambda}u}{[\lambda]}
  & q^{1-r}\,\sqrt{\frac{[\lambda+1][\lambda-1]}{[\lambda]^2}} \\
  q^{1-r}u\,\sqrt{\frac{[\lambda+1][\lambda-1]}{[\lambda]^2}}
  & \frac{(-1)^\lambda q^{1-\lambdab}}{[\lambda]}
  \end{array}\right) 
\end{equation}
in the basis $(A_+,A_-)$,
where $u$ is an undetermined complex number of modulus one. 
Since, by construction, the matrix elements of $\F_{A[2,1]A}^{[2,1]}$ are rational functions of $(q,q^{-1})$, this form holds for any $q\in\Cbb^\times$.
Then, we assume that the rows and columns of $\F_{A[2,1]A}^{[2,1]}$ can be rescaled with an appropriate gauge transformation \eqref{eq:gauge.F}, which yields the $U_q(sl_2)$ associativity matrix \eqref{eq:F2}:
\begin{equation}
  \F_{A[2,1]A}^{[2,1]} \to \left(\begin{array}{cc}
  \frac{(-1)^{\lambda}}{[\lambda]}
  & \sqrt{\frac{[\lambda+1][\lambda-1]}{[\lambda]^2}} \\
  \sqrt{\frac{[\lambda+1][\lambda-1]}{[\lambda]^2}}
  & \frac{(-1)^{\lambda+1}}{[\lambda]}
  \end{array}\right) = F_{\lambda 2\lambda}^2 \,,
  \qquad A=[\lambda,\lambdab] \,.
\end{equation}

\paragraph{Example 2: $D=[3,1]$.}
We set $A=[\lambda,\lambdab]$, $B=[2,1]$, $C=[\mu,\mub]$ and $D=[3,1]$. There are two cases where the space $\mathrm{Hom}(D,A\otimes B\otimes C)$ is non-zero: it is two-dimensional for $|\mu-\lambda|=1$ and $\mub=\lambdab$, and it is one-dimensional for $|\mu-\lambda|=3$ and $\mub=\lambdab$.
Let us study the case
$$A=Q_{k,-q^r}=[\lambda,\lambdab] \quad B=[2,1] \quad C=A_+=[\lambda+1,\lambdab] \,, \quad D=[3,1] \,.$$
with $0\leq k<r$.
We set $X=C_-$, and $Y\in\{A_+,A_-\}$.
Using computer algebra, we obtain the normalisation factors (\ref{eq:Z_ABC}--\ref{eq:X_ABC})
\begin{equation}
  \Omega_{[31][\lambda\lambdab][\lambda+2,\lambdab]}=\frac{1}{[\lambda+2][\lambdab]}
  \qquad \Omega_{[31][\lambda\lambdab][\lambda,\lambdab]}=\frac{[2][\lambda-1]}{[\lambda+1][\lambda][\lambdab]}
  \qquad \Omega_{[31][\lambda\lambdab][\lambda-2,\lambdab]}=\frac{1}{[\lambda][\lambdab]}
\end{equation}
and the overlaps \eqref{eq:Z_YX}
\begin{equation}
  \left(Z_{A[2,1]A_+}^{[3,1]} \right)_{A_+,A} 
  = q^{1-r} \,,
  \qquad \left(Z_{A[2,1]A_+}^{[3,1]} \right)_{A_-,A} 
  = \frac{(-1)^\lambda q^{1/2-\lambdab}}{[\lambda]} \,.
\end{equation}
Thus we get
\begin{equation}
  \left(\F_{A[2,1]A_+}^{[3,1]} \right)_{A_+,A}
  = q^{1-r} \sqrt{\frac{[\lambda-1][\lambda+2]}{[\lambda][\lambda+1]}} \,,
  \quad \left(\F_{A[2,1]A_+}^{[3,1]} \right)_{A_-,A} 
  = (-1)^\lambda q^{1/2-\lambdab} \sqrt{\frac{[2]}{[\lambda][\lambda+1]}} \,.
\end{equation}
Arguing on unitarity like in the previous case, we obtain a matrix of the form
\begin{equation}
  \F_{A[2,1]A_+}^{[3,1]} = \left(\begin{array}{cc}
    (-1)^{\lambda+1} q^{3/2-\lambda}u \sqrt{\frac{[2]}{[\lambda][\lambda+1]}}
    & q^{1-r} \sqrt{\frac{[\lambda-1][\lambda+2]}{[\lambda][\lambda+1]}} \\
    q^{1-r}u \sqrt{\frac{[\lambda-1][\lambda+2]}{[\lambda][\lambda+1]}}
    & (-1)^\lambda q^{1/2-\lambdab} \sqrt{\frac{[2]}{[\lambda][\lambda+1]}}
  \end{array}\right)
\end{equation}
in the basis $([\lambda+2,\lambdab],A)$ for $X$ and $(A_+,A_-)$ for $Y$.
By rescaling the rows and columns with \eqref{eq:gauge.F}, we get
the $U_q(sl_2)$ associativity matrix:
\begin{equation}
  \F_{[\lambda,\lambdab][2,1][\lambda+1,\lambdab]}^{[3,1]} \to \left(\begin{array}{cc}
    (-1)^\lambda \sqrt{\frac{[2]}{[\lambda][\lambda+1]}}
    & \sqrt{\frac{[\lambda-1][\lambda+2]}{[\lambda][\lambda+1]}} \\
    \sqrt{\frac{[\lambda-1][\lambda+2]}{[\lambda][\lambda+1]}}
    & (-1)^{\lambda+1} \sqrt{\frac{[2]}{[\lambda][\lambda+1]}}
  \end{array}\right)
  = F_{\lambda, 2, \lambda+1}^3 \,.
\end{equation}
Similary results hold for $\F_{[\lambda,\lambdab][2,1][\lambda-1,\lambdab]}^{[3,1]}$.

\paragraph{Example 3: $D=[1,2]$.}
We set $A=[\lambda,\lambdab]$, $B=[2,1]$, $C=[\mu,\mub]$ and $D=[1,2]$. The space $\mathrm{Hom}(D,A\otimes B\otimes C)$ is one-dimensional if $|\lambda-\mu|=|\lambdab-\mub|=1$, and zero otherwise.
We can take, for instance, $C=[\lambda+1,\lambdab+1]$, which yields $X=[\lambda,\lambdab+1]$ and $Y=[\lambda+1,\lambdab]$. We then obtain
\begin{equation} \label{eq:F[12]}
  \F_{[\lambda,\lambdab][21][\lambda+1,\lambdab+1]}^{[12]} = Z_{[\lambda,\lambdab][21][\lambda+1,\lambdab+1]}^{[12]} = q^{1-r} \,.
\end{equation}
Of course, $\F_{A[21]C}^{[12]}$ can be rescaled to any value using \eqref{eq:gauge.F}, in particular 
to $F_{\lambda,2,\lambda+1}^1=1$. The significance of \eqref{eq:F[12]} is that $\F_{A[21]C}^{[12]}$ is unitary when $|q|=1$.

\paragraph{Example 4: $D=[2,2]$.}
We set $A=[\lambda,\lambdab]$, $B=[2,1]$, $C=[\mu,\mub]$ and $D=[2,2]$. There are two cases where the space $\mathrm{Hom}(D,A\otimes B\otimes C)$ is non-zero: it is two-dimensional for $\mu=\lambda$ and $|\mub-\lambdab|=1$, and it is one-dimensional for $|\mu-\lambda|=2$ and $|\mub-\lambdab|=1$.
Let us study the case
$$A=Q_{k,-q^r}=[\lambda,\lambdab]\,, \quad B=[2,1]\,, \quad C=[\lambda,\lambdab+1] \,, \quad D=[2,2] \,.$$
Using computer algebra, we obtain the normalisation factors
\begin{equation}
  \Omega_{[22][\lambda,\lambdab][\lambda+1,\lambdab+1]} = \frac{1}{[2][\lambda+1][\lambdab]} \,,
  \qquad \Omega_{[22][\lambda,\lambdab][\lambda-1,\lambdab+1]}
  = \frac{-1}{(q-q^{-1})^2[2][\lambda][\lambdab]} \,,
\end{equation}
and the overlaps \eqref{eq:Z_YX} with $X=C_-=[\lambda-1,\lambdab+1]$
\begin{equation}
  \left(Z_{A[21]C}^{[22]} \right)_{A_+,C_-} = q^{-1/2} \,,
  \qquad \left(Z_{A[21]C}^{[22]} \right)_{A_-,C_-} = \frac{(-1)^{r+k+1}q^k(q-q^{-1})}{[r+k]} \,,
\end{equation}
which yields
\begin{equation}
  \left(\F_{A[21]C}^{[22]} \right)_{A_+,C_-} = q^{-1/2} \sqrt{\frac{[\lambda+1][\lambda-1]}{[\lambda]^2}} \,,
  \qquad \left(\F_{A[21]C}^{[22]}\right)_{A_-,C_-}
  = \frac{iq^k (-1)^{\lambda+1}}{[\lambda]} \,.
\end{equation}
The unitarity argument yields an associativity matrix of the form
\begin{equation}
  \F_{[\lambda,\lambdab][2,1][\lambda,\lambdab+1]}^{[2,2]} = \left(\begin{array}{cc}
  \frac{iq^{-k-1}(-1)^{\lambda+1}u}{[\lambda]}
  & q^{-1/2}\,\sqrt{\frac{[\lambda+1][\lambda-1]}{[\lambda]^2}} \\
  q^{-1/2}u\,\sqrt{\frac{[\lambda+1][\lambda-1]}{[\lambda]^2}}
  & \frac{iq^k (-1)^{\lambda+1}}{[\lambda]}
  \end{array}\right) \,.
\end{equation}
After rescaling the rows and columns by a gauge transformation \eqref{eq:gauge.F}, we obtain
\begin{equation}
  \F_{[\lambda,\lambdab][2,1][\lambda,\lambdab+1]}^{[2,2]} \to \left(\begin{array}{cc}
  \frac{(-1)^{\lambda}}{[\lambda]}
  & \sqrt{\frac{[\lambda+1][\lambda-1]}{[\lambda]^2}} \\
  \sqrt{\frac{[\lambda+1][\lambda-1]}{[\lambda]^2}}
  & \frac{(-1)^{\lambda+1}}{[\lambda]}
  \end{array}\right) = F_{\lambda 2 \lambda}^2 \,.
\end{equation}

\section{Conclusion}
\label{sec:conclusion}

To conclude, let us describe some consequences of our results on the modules of $\TL^a$, in terms of Conformal Field Theory (CFT).
\medskip

Consider the dense loop model on the square lattice, with loop weight $\beta=-q-q^{-1}$. In the regime $\beta\in [-2,2]$, this model is critical, and scales to a CFT with central charge $c=1-6(1-g)^2/g$, where 
$g=1-\tfrac{1}{\pi}\mathrm{Arcos}(\beta/2)$. We thus set $q=\exp(i\pi g)$, with $0\leq g \leq 1$.
We suppose that $q$ is not a root of unity, and hence the central charge is generic.
The module $[\lambda,\lambdab]$ scales as
\begin{equation} \label{eq:lim.Q}
  [\lambda,\lambdab] \to \bigoplus_{r=1}^{\infty} (K_{r,\lambda},K_{r,\lambdab})
\end{equation}
where $K_{r,s}$ denotes the simple quotient of the Verma module with conformal weight $h_{rs}$, and $(K,\Kb)$ is the Virasoro module with holomorphic part $K$ and anti-holomorphic part $\Kb$.
This limit stems from the correspondence between the loop model and the six-vertex model \cite{Baxter82}, and the convergence of the latter to the free scalar CFT \cite{Nienhuis84,DiF87}. The meaning of \eqref{eq:lim.Q} is the following: if we denote by $\Lambda_0(N),\Lambda_1(N),\Lambda_2(N)\dots$ the transfer-matrix eigenvalues in decreasing absolute value, then as $N\to \infty$, the eigenstates of $[\lambda,\lambdab]$ with a gap $\log[\Lambda(N)/\Lambda_0(N)]$ of order $(-1/N)$ converge to the states of the RHS of \eqref{eq:lim.Q}.
\medskip

Since $\cal Q$ is a semi-simple ribbon category, we can expect its scaling limit to be a bulk CFT which we call $\cal C$, with a well-defined Operator Product Expansion (OPE), and conformal blocks on the Riemann sphere satisfying the algebraic axioms of \cite{MooreSieb89}. In the theory $\cal C$, the primary operators have dimensions $(h,\hb)=(h_{r,\lambda},h_{r,\lambdab})$ for any positive integers $r,\lambda,\lambdab$, and each of these operators (which we denote as $\Phi_{r|\lambda,\lambdab}$) generates a simple $\mathsf{Vir}\otimes\overline{\mathsf{Vir}}$ module. This CFT $\cal C$ is similar to the generalised minimal models of \cite{Zam05}, except that it admits some non-scalar primary operators, corresponding to $\lambda\neq \lambdab$.
\medskip

All the states of $(K_{r,\lambda},\Kb_{r,\lambdab})$ have conformal spin of the form $(h_{r\lambda}-h_{r\lambdab})+\Zbb$, where
\begin{equation}
  h_{rs} = \frac{(r-sg)^2-(1-g)^2}{4g} \,.
\end{equation}
Hence, under a rotation of angle $2\pi$, these states are multiplied by a factor
\begin{equation}
  \exp[2i\pi(h_{r\lambda}-h_{r\lambdab})] = (-1)^{r(\lambda-\lambdab)} \exp\left[\frac{i\pi g}{2}(\lambda^2-\lambdab^2) \right] \,.
\end{equation}
On the lattice, this rotation is implemented by the twist operator $\theta=\Omega^N$, with eigenvalue $(-q^{1/2})^{\lambda^2-\lambdab^2}$ on $[\lambda,\lambdab]$. We thus get the scaling
\begin{equation}
  \theta \to (-1)^{(r+1)(\lambda-\lambdab)} \, \exp[2i\pi(L_0-\Lb_0)]
\end{equation}
in the component $(K_{r,\lambda},\Kb_{r,\lambdab})$ of $[\lambda,\lambdab]$.
\medskip

Furthermore, from the decomposition \eqref{eq:decomp.QxQ}, we expect the OPE in $\cal C$ to read
\begin{equation}
  \Phi_{r|\lambda,\lambdab} \cdot \Phi_{s|\mu,\mub} \to
  \sum_{t,\nu,\nub} n_{rs}^t \, n_{\lambda\mu}^\nu  \, n_{\lambdab\mub}^{\nub} \,  \Phi_{t|\nu,\nub} \,,
\end{equation}
where $\{n_{ij}^k\}$ are the $sl_2$ fusion numbers given in \eqref{eq:n}. This OPE is consistent with the Virasoro fusion rules on the holomorphic and anti-holomorphic parts. The crossing symmetry of the four-point conformal blocks on the sphere is encoded in the braiding matrices $\sigma_{AB}^C$ and the associativity matrices $\F_{ABC}^D$.
\medskip

With the above remarks, we have only sketched the analysis of the CFT $\cal C$ describing the scaling limit of $\cal Q$, which is a specific sector of the critical loop model. We believe that a more thorough study of this CFT will be of interest for the understanding of the relations between non-rational CFTs and correlations in loop models.

\appendix
\section*{Appendix}

\section{Irreducible standard modules}
\label{app:std}

In this section, we consider the irreducible standard modules, i.e. the modules $W_{k,x}$ where $x$ is not of the form $\pm q^r$ with $r\in k+\Zbb$ and $|r|>|k|$.
\medskip

We describe the tensor products of the form $W_{k,x}\otimes W_{\ell,y}$ and $W_{k,x}\otimes [\mu,\mub]$, where $W_{k,x}$ and $W_{\ell,y}$ are irreducible, and $[\mu,\mub]\in\cal Q$, in the
case when $q$ is not a root of unity. These ``fused'' modules are studied from a more mathematical point of view in \cite{IMD-unpub}.
Here, we show that $(W_{k,x}\otimes W_{\ell,y})(N)$ is infinite-dimensional for any $N$, and that the following isomorphism holds
\begin{equation} \label{eq:WxW/F=W}
  (W_{k,x}\otimes W_{\ell,y}) \Big/ \aver{F=f_{m,z} \,, \ \Fb=\fb_{m,z}}
  \equiv W_{m,z}
\end{equation}
for any $W_{m,z}$ such that $z\notin \pm q^{m+\Zbb}$. Here, $f_{m,z}=q^m z+ q^{-m}z^{-1}$, and $\fb_{m,z}=f_{-m,z}$ are the values of the braid matrices $F,\Fb$ on $W_{m,z}$.
\medskip

Furthermore, using the structure of $W_{k,x}\otimes W_{\ell,y}$ and the properties of the Gram form, we argue that an irreducible standard module cannot admit a left or a right dual in the sense of monoidal categories.
\medskip

These results are valid when $q$ is not a root of unity.

\subsection{Tensor products}
\label{app:fusion}

\paragraph{The tensor product $W_{k,x}\otimes W_{\ell,y}$.}
Let $k,\ell\in \Zbb_{\geq 0}/2$ and $x,y\in\Cbb^\times$, and consider the modules $A=W_{k,x}$ and $B=W_{\ell,y}$. For simplicity, we assume that $(A,B)\neq (W_{0,i},W_{0,i})$.
To describe the structure of $W_{k,x}\otimes W_{\ell,y}$, we introduce the states $u_{k,\ell}^m \in (A\otimes B)(2m)$ for any $m\in \Zbb_{\geq 0}/2$ such that $m+k+\ell$ is an integer:
\begin{equation} \label{eq:u_kl^m}
  u_{k,\ell}^m := \begin{cases}
    (c_0^{\dag(m-k-\ell)}u_k ) \otimes u_\ell & \text{if } m\geq k+\ell \,, \\
    c_0^{k+\ell-m}\cdot (u_k\otimes u_\ell)
    & \text{if } m\leq k+\ell \,.
  \end{cases}
\end{equation}
The states $u_{k,\ell}^m$ satisfy\footnote{This is where the case $(A,B)=(W_{0,i},W_{0,i})$ is different, and should be discussed separately.} $u_{k,\ell}^m\propto c_0\cdot u_{k,\ell}^{m+1}$.
Moreover, the space $(A\otimes B)(N)$ is spanned by states of the form
\begin{equation}  \label{eq:basis.WxW}
  \begin{aligned}
    & \cd_{i_1}\dots \cd_{i_{N/2}} f^j \cdot u_{k,\ell}^0 \,,
    && j\in \Zbb_{\geq 0} \,, && \text{if } k+\ell \in\Zbb \,, \\
    & \cd_{i_1}\dots \cd_{i_{N/2-m}} \Omega^j \cdot u_{k,\ell}^m \,,
    &\quad& j\in \Zbb \,, &\quad&  0<m\leq N/2 \,.
  \end{aligned}
\end{equation}
Therefore, the submodules defined as
\begin{equation} \label{eq:Xm}
  X_m(N):=\TL^a(N,2m)\cdot u_{k,\ell}^m
\end{equation}
form a filtration of $(A\otimes B)(N)$:
\begin{equation} 
  X_{m_0}(N)
  \subseteq X_{m_0+1} \subseteq \dots \subseteq X_{N/2}(N) = (A\otimes B)(N) \,,
\end{equation}
where $m_0=0$ if $k+\ell$ is an integer, and $m_0=\tfrac12$ if 
$k+\ell$ is a half-integer.
\medskip

We admit that the states \eqref{eq:basis.WxW} are linearly independent, and hence $(A\otimes B)(N)$ is infinite-dimensional.

\paragraph{The module $W_{m,z}$ as a quotient of $(W_{k,x}\otimes W_{\ell,y})$.}
Let us prove the statement \eqref{eq:WxW/F=W}.
Consider the space
\begin{equation}
  Z(N):=\mathrm{span}\big[
      (F-f_{mz}\,\id)\cdot u + (\Fb-\fb_{mz}\,\id)\cdot v\,,
      \quad u,v\in (A\otimes B)(N)
      \big] \,.
\end{equation}
Since $F,\Fb$ are central elements, $Z$ is a submodule of $A \otimes B$.
We thus denote
\begin{equation}
  X(N):= (A\otimes B)(N) \,, \qquad \wh{X}(N):= X(N) \big/ Z(N) \,.
\end{equation}
In these notations, the relation \eqref{eq:WxW/F=W} reads $\wh{X} \equiv W_{m,z}$.
Recall that the submodules $X_s(N)=\TL^a(N,2s)\cdot u_{k\ell}^s$ form a filtration \eqref{eq:filtration} for $A\otimes B$.
We shall also use the identities
\begin{equation}
  \begin{aligned}
    &F\cdot u_{k\ell}^0=\Fb\cdot u_{k\ell}^0= f\cdot u_{k\ell}^0 \,, \\
    &F\cdot u_{k\ell}^s\equiv (q^s\Omega+q^{-s}\Omega^{-1})\cdot u_{k\ell}^s \mod X_{s-1} \qquad \text{if } s>0 \,, \\
    & \Fb\cdot u_{k\ell}^s\equiv (q^{-s}\Omega+q^s\Omega^{-1})\cdot u_{k\ell}^s \mod X_{s-1} \qquad \text{if } s>0 \,.
  \end{aligned}
\end{equation}
For any $u\in X(N)$, we denote by $\wh{u}$ the equivalence class of $u$ modulo $Z(N)$. We define the submodules $\wh{X}_s(N):=\TL^a(N,2s)\cdot \wh{u}_{k\ell}^s$.
Like in $X(N)$, the submodules $\wh{X}_s(N)$ yield a filtration of $\wh{X}(N)$.
We then have
\begin{equation}
  \begin{aligned}
    & f\cdot\wh{u}_{k\ell}^0= f_{mz}\,\wh{u}_{k\ell}^0 = \fb_{mz}\,\wh{u}_{k\ell}^0 \,, \\
    & f_{mz}\,\wh{u}_{k\ell}^s=F\cdot \wh{u}_{k\ell}^s\equiv (q^s\Omega+q^{-s}\Omega^{-1})\cdot\wh{u}_{k\ell}^s \mod \wh{X}_{s-1} \qquad \text{if } s>0\,, \\
    & \fb_{mz}\,\wh{u}_{k\ell}^s=\Fb\cdot \wh{u}_{k\ell}^s \equiv (q^{-s}\Omega+q^s\Omega^{-1})\cdot\wh{u}_{k\ell}^s \mod \wh{X}_{s-1}  \qquad \text{if }  s>0\,.
  \end{aligned}
\end{equation}
By combining linearly the above relations, we obtain
\begin{equation} \label{eq:f.u.and.Om.u}
  \begin{aligned}
    & f\cdot\wh{u}_{k\ell}^0= f_{mz}\,\wh{u}_{k\ell}^0 = \fb_{mz}\,\wh{u}_{k\ell}^0 \,, \\
    & \Omega\cdot \wh{u}_{k\ell}^s\equiv \frac{q^s f_{mz}-q^{-s}\fb_{mz}}{q^{2s}-q^{-2s}} \, \wh{u}_{k\ell}^s \mod \wh{X}_{s-1}  && \text{if } s>0 \,, \\
    & \Omega^{-1}\cdot \wh{u}_{k\ell}^s\equiv \frac{q^s \fb_{mz}-q^{-s}f_{mz}}{q^{2s}-q^{-2s}} \, \wh{u}_{k\ell}^s \mod \wh{X}_{s-1}   \qquad && \text{if } s>0 \,.
  \end{aligned}
\end{equation}
Thus, for any $s\in\{m_0,\dots,N/2\}$, we have
\begin{equation}
  \big[(q^s f_{mz}-q^{-s}\fb_{mz})(q^s \fb_{mz}-q^{-s}f_{mz})
    - (q^{2s}-q^{-2s})^2\big]
  \, \wh{u}_{k\ell}^s\equiv 0 \mod \wh{X}_{s-1} \,,
\end{equation}
which can be factorised as
\begin{equation}
  [s-m]\, [s+m]\, (z^2-q^{2s})(z^2-q^{-2s}) \, \wh{u}_{k\ell}^s\equiv 0 \mod \wh{X}_{s-1} \,.
\end{equation}
Since, by hypothesis, $z$ is not of the form $\pm q^n$ with $n \in m_0/2+\Zbb$, we obtain $\wh{u}_{k\ell}^s\equiv 0 \mod \wh{X}_{s-1}$ for any $s\neq m$.
This yields, for any $N$,
\begin{equation}
  \begin{aligned}
    &\wh{X}(N)=\wh{X}_{N/2}(N) \subseteq \wh{X}_{N/2-1}(N) \subseteq\dots \subseteq \wh{X}_{m}(N) \,, \\
    &\wh{X}_{m-1}(N) \subseteq \wh{X}_{m-2}(N) \subseteq\dots \subseteq \wh{X}_{m_0(k,\ell)-1}(N)=0 \,,
  \end{aligned}
\end{equation}
and hence 
\begin{equation}
  \wh{X}(N)=\wh{X}_m(N)=\TL^a(N,2m)\cdot \wh{u}_{k\ell}^m 
  \qquad \text{and}\qquad \wh{X}_{m-1}(N)=0 \,.
\end{equation}
For $s=m$, \eqref{eq:f.u.and.Om.u} yields
\begin{equation}
  \Omega\cdot \wh{u}_{k\ell}^m \equiv z\, \wh{u}_{k\ell}^m \mod \wh{X}_{m-1}
  \qquad  \text{if } m>0 \,.
\end{equation}
Since $\wh{X}_{m-1}=0$ and $c_1\cdot \wh{u}_{k\ell}^m\in\wh{X}_{m-1}(2m-2)$, we obtain
\begin{equation} \label{eq:uh(2m).is.seed}
  \begin{aligned}
    & f\cdot \wh{u}_{k\ell}^0 = (z+z^{-1}) \, \wh{u}_{k\ell}^0
    &\qquad& \text{if } m=0 \,, \\
    &\Omega\cdot \wh{u}_{k\ell}^m = z\, \wh{u}_{k\ell}^m
    && \text{if } m>0 \\
    & c_1\cdot \wh{u}_{k\ell}^m = 0
    && \text{if } m\geq 1 \,.
  \end{aligned}
\end{equation}
Let us show that the state $\wh{u}_{k\ell}^m$ is non-zero.
If $m>0$, any element of $X(2m)$ is of the form $u\equiv\lambda\, u_{k\ell}^m \mod X_{m-1}$, where $\lambda$ is a Laurent polynomial in $\Omega$.
Moreover, we have
\begin{equation}
  \begin{aligned}
    &(F-f_{mz}\,\id)\cdot u_{k\ell}^m
    \equiv (q^m\,\id-q^{-m}z^{-1}\Omega^{-1})(\Omega-z\,\id)\cdot u_{k\ell}^m \mod X_{m-1} \,,\\
    &(\Fb-\fb_{mz}\,\id)\cdot u_{k\ell}^m
    \equiv (q^{-m}\,\id-q^m z^{-1}\Omega^{-1})(\Omega-z\,\id)\cdot u_{k\ell}^m \mod X_{m-1} \,.
  \end{aligned}
\end{equation}
Hence, any element of $Z(2m)$ is of the form $v\equiv\lambda\,(\Omega-z\,\id)\cdot u_{k\ell}(2m)\mod X_{m-1}$, where $\lambda$ is a Laurent polynomial in $\Omega$. This shows that $u_{k\ell}^m\notin Z(2m)$, and hence $\wh{u}_{k\ell}^m\neq 0$.
Similarly, if $m=0$, any element of $X(0)$ is of the form $u=\lambda\, u_{k\ell}^0$, and any element of $Z(0)$ is of the form $v=\lambda (f-f_{0z}\,\id)u_{k\ell}^0$, where $\lambda$ is a polynomial in $f$ in both cases. With the same argument as for $m>0$, we then conclude that $\wh{u}_{k\ell}^0\neq 0$.
\medskip

Finally, we have shown that $\wh{X}(N)=\TL^a(N,2m)\cdot \wh{u}_{k\ell}^m$ and $\wh{u}_{k\ell}^m$ is a seed state of type $(m,z)$. Since $W_{m,z}$ is simple, this proves \eqref{eq:WxW/F=W}.

\paragraph{The tensor product $W_{k,x} \otimes [\mu,\mub]$.}
Let $W_{k,x}$ be an irreducible standard module, and $[\mu,\mub]$ a simple module in $\cal Q$.
Through an argument similar to the case of $[\lambda,\lambdab]\otimes [\mu,\mub]$, one obtains the decomposition
\begin{equation}
  W_{k,x} \otimes [\mu,\mub] \equiv
  \bigoplus_{i=-\frac{\mu-1}2}^{\frac{\mu-1}2}
  \bigoplus_{j=-\frac{\mub-1}2}^{\frac{\mub-1}2} W_{k+i-j,xq^{i+j}} \,.
\end{equation}

\subsection{Absence of duality}

\paragraph{Definitions.}
In a monoidal category $\cal M$ with unit element $V$, we say that an object $A$ admits a left dual $A^*$ iff $A^*\in \cal M$ and there exists a pair of morphisms
$$\eps_A:A^*\otimes A\to V \qquad \text{and} \qquad \eta_A:V\to A\otimes A^*$$
called respectively the evaluator and coevaluator, which satisfy the relations
\begin{equation} \label{eq:left.dual}
  (\id_A\otimes \eps_A) \cdot (\eta_A \otimes \id_A) = \id_A \,,
  \qquad (\eps_A \otimes \id_{A^*}) \cdot (\id_{A^*} \otimes \eta_A) = \id_{A^*} \,.
\end{equation}
The right dual is defined in a similar way, with morphisms $\bar\eps_A:A\otimes A^*\to V$ and $\bar\eta_A:V\to A^*\otimes A$.

\paragraph{Bilinear form associated to $\eps_A$.}
Let $A\in \cal M$, and assume that $A$ admits a left dual $A^*$.
Let $\xi=\eta_A(v_0)$. Any $v\in V(N)$ can be written as $v=\lambda\cdot v_0$ with $\lambda\in \TL^a(N,0)$, and thus $\eta_A(v)=\eta_A(\lambda\cdot v_0)=\lambda\cdot \xi$. Hence, $\xi\neq 0$, otherwise $\eta_A$ would be identically zero. Moreover, $f\cdot \xi = f\cdot \eta_A(v_0)=\eta_A(f\cdot v_0) = \beta\, \xi$, and therefore $\xi$ is a seed state of type $(0,-q)$ in $A\otimes A^*$, and $\eta_A$ is the corresponding insertion map $V\to A\otimes A^*$.
Let us now examine the morphism $\eps_A$. Since the Gram form $\aver{\,,\,}$ on $V\times V$ is non-degenerate, by combining  \eqref{eq:F=F'}, \eqref{eq:Hom=F} and \eqref{eq:Kl}, we obtain the isomorphism of vector spaces
\begin{equation}
  \mathrm{Hom}(A^*\otimes A,V) \equiv \I(A^*\otimes A,V)
  \equiv \I(A,\rho(A^*)\otimes V) \equiv \I(A,\rho(A^*)) \,.
\end{equation}
More specifically, there exists a unique bilinear form $g_A\in \I(A,\rho(A^*))$ such that
\begin{equation} \label{eq:eps}
  \eps_A(u\otimes v) = \sum_i g_A[v,c_1\dots c_L \cdot r_A(\rho(u)\otimes \vthetab_i)] \, \vtheta_i \,,
\end{equation}
for any $u\in A^*(L)$ and $v\in A(M)$, where $(\vtheta_i),(\vthetab_i)$ is any pair of mutually dual bases of $V(L+M)$.
For any $u\in A^*(L)$ and $v\in \overline{\rm Rad}(g_A,M)$, we have $\eps_A(u\otimes v)=0$. Hence, the LHS of the first equation in \eqref{eq:left.dual} vanishes when restricted to $\overline{\rm Rad}(g_A)$: this proves that $\overline{\rm Rad}(g_A)=0$. A similar argument yields $\mathrm{Rad}(g_A)=0$, and thus $g_A$ is non-degenerate. In particular, we obtain $\dim A^*(N)=\dim A(N)$ for any $N$.

\paragraph{Simple modules.}
Let $A$ and $B$ be two simple $\TL^a$ modules. The space of invariant bilinear forms on $A\times B$ is given by
\begin{equation} \label{eq:gram.is.unique}
  \I(A,B) = \begin{cases}
    \mathrm{span} \aver{\,,\,} & \text{if } A=\rho(B) \,, \\
    0 & \text{otherwise.}
  \end{cases}
\end{equation}
Let $A$ be a simple module $A$, and suppose that it admits a left dual $A^*$. Then, by the above argument, $g_A$ is a non-degenerate form on $A\times \rho(A^*)$. Let $B$ be a nonzero irreducible submodule of $\rho(A^*)$, and $h_A\in\I(A,B)$ be the restriction of $g_A$ to $A\times B$. Since $g_A$ is non-degenerate, we have $h_A\neq 0$, which yields $B\equiv \rho(A)$ by \eqref{eq:gram.is.unique}. Using the property $\dim A(N)=\dim A^*(N)$ we obtain  $B=\rho(A^*)$, and finally $A^*\equiv A$.
\medskip

If $A=W_{k,x}$ is an irreducible standard module, then we assume that the states $f^j u_{k,k}^0$ with integer $j\geq 0$ form a basis of $(A\otimes A)(0)$ (see Sec.\ref{app:fusion}), and hence there cannot exist an eigenvector of $f$ in $(A\otimes A)(0)$. This contradicts the fact that $\xi=\eta_A(v_0)$ is a seed state of type $(0,-q)$ in $A\otimes A$.
\medskip

Therefore, an irreducible standard module never admits a left dual. One can argue similarly about right duals.

\section{The bilinear form $\aaver{\,,\,}$}
\label{app:form}

Consider the case when $q$ is not a root of unity, and let
$$A=Q_{k,x} \,, \qquad B=Q_{\ell,y} \,, \qquad C=Q_{m,z}$$
be three simple modules in $\cal Q$, such that $N_{AB}^C=1$.
In Sec.~\ref{sec:decomp}, we have shown that there exists a unique invariant bilinear form $\aaver{\,,\,}$ on $(A\otimes B)\times \Cb$ such that $\aaver{w_{k\ell}^m,\ub_m}=1$.
Here, we develop a simple algorithm to compute $\aaver{u,w}$ for any $u\in (A\otimes B)(N)$ and $w\in \Cb(N)$.
\medskip

The basic ingredient is to write $c_j\cdot w_{k\ell}^m$ in terms of $w_{k,\ell}^{m'}$ with $m'<m$.
Let $j=0,\dots,2m-1$, and $n:=m-k-\ell$.
There are three cases to consider:
\begin{subequations} \label{eq:cj.uklm}
\begin{itemize}
\item If $m>k+\ell$ then
  \begin{equation}
    c_j \cdot w_{k,\ell}^m = \begin{cases}
      \gamma(\ell,y^{-1})\, w_{k,\ell}^{m-1} & \text{if } j=0 \\
      c^\dag_{2k+2n-j-2} \cdot w_{k,\ell}^{m-2} & \text{if } 1\leq j\leq n-1 \\
      \gamma(k,x)\, w_{k,\ell}^{m-1} & \text{if } j=n \\
      \gamma(k,x^{-1})\, w_{k,\ell}^{m-1} & \text{if } j=2k+n \\
      c^\dag_{2k+2n-j} \cdot w_{k,\ell}^{m-2} & \text{if } 2k+n+1\leq j\leq 2k+2n-1 \\
      y\, \Omega^{-1}\cdot w_{k,\ell}^{m-1} & \text{if } j=2k+2n \\
      0 & \text{otherwise,}
    \end{cases}
  \end{equation}
  where $\gamma(k,x)=x$ if $k>0$ and $\gamma(0,x)=x+x^{-1}$.
\item If $|k-\ell|< m \leq k+\ell$ then
  \begin{equation}
    c_j \cdot w_{k,\ell}^m = \begin{cases}
      w_{k,\ell}^{m-1} & \text{if } j=0 \\
      x^{-1}y \, w_{k,\ell}^{m-1} & \text{if } j=2k+n \\
      0 & \text{otherwise.}
    \end{cases}
  \end{equation}
  \item If $0<m\leq |k-\ell|$ then
  \begin{equation}
    c_j \cdot w_{k,\ell}^m = \begin{cases}
      w_{k,\ell}^{m-1} & \text{if } j=0 \\
      0 & \text{otherwise.}
    \end{cases}
  \end{equation}
\end{itemize}
\end{subequations}
The relations \eqref{eq:cj.uklm} can be written
\begin{equation}
  c_j \cdot w_{k,\ell}^m = \mu\cdot w_{k,\ell}^{m'} \,,
  \qquad \lambda \in \TL^a(2m-2,2m') \,, \quad m'<m \,.
\end{equation}
Let $u\in (A\otimes B)(N)$ and $w\in \Cb(N)$ be two link states, and let us write $N=2p$. The state $u$ can be written $u=\lambda\cdot w_{k,\ell}^p$, and thus
\begin{equation}
  \aaver{u,w} = \aaver{\mu\,w_{k\ell}^p,w} = \aaver{w_{k\ell}^p,w'} \,,
  \qquad w'=\mu^t\,w \,.
\end{equation}
The calculation of $\aaver{\,,\,}$ thus reduces to that of $\aaver{w_{k\ell}^p,w}$, for any link state $w\in\Cb(2p)$. This is done with the following recursive algorithm.
\begin{itemize}
\item If $p<m$, then $w=0$ and hence $\aaver{w_{k\ell}^p,w}=0$.
\item If $p=m$, then $w=\ub_m$, and hence $\aaver{w_{k\ell}^m,\ub_m}=1$.
\item If $p>m$, then there exists $j\in\{0,\dots,2p-1\}$ and $w'\in \Cb(2p-2)$ such that $w=c_j^\dag\cdot w'$, and we can write
\begin{equation}
  \aaver{w_{k\ell}^p,w} = \aaver{w_{k\ell}^p,c_j^\dag\cdot w'}
  = \aaver{c_j\cdot w_{k\ell}^p,w'}
  = \aaver{\lambda\cdot w_{k\ell}^{p'},w'}
  = \aaver{w_{k\ell}^{p'},\lambda^\dag\cdot w'}
\end{equation}
with $\lambda \in \TL^a(2p-2,2p')$ and $p'<p$ obtained from \eqref{eq:cj.uklm}. We thus obtain $\aaver{w_{k\ell}^p,w} = \aaver{w_{k\ell}^{p'},w''}$, where $w''=\lambda^\dag\cdot w'$ is a link state of $\Cb(2p')$, with $p'<p$.
\end{itemize}

Moreover, using the isomorphisms \eqref{eq:Kr} and \eqref{eq:Kl}, there exists two non-zero constants $a_{ABC}$ and $b_{ABC}$ such that
\begin{equation} \label{eq:turn.form}
  \begin{aligned}
    & \aaver{\vb,c_1\dots c_L\cdot(u\otimes w)}
    = a_{ABC}\, \aaver{\ub\otimes \vb,w} \,, \\
    & \aaver{\ub,c_{L+1}\dots c_{L+M} \cdot(w\otimes v)}
    = b_{ABC}\, \aaver{\ub\otimes \vb,w} \,,
  \end{aligned}
\end{equation}
for any $u\in A(L)$, $v\in B(M)$ and $w\in C(L+M)$. Here, we use the short-hand notations $\ub=\rho(u)$ and $\vb=\rho(v)$.
Using the fact that $\aaver{w_{k\ell}^m,\ub_m}=1$ in $(A\otimes B)\times \Cb$, the constants $a_{ABC}$ and $b_{ABC}$ can be expressed as
\begin{equation}
  a_{ABC} = \begin{cases}
    \aaver{\Omega^{-2k} c_0^{\dag(k+m-\ell)}\ub_\ell,w_{k,m}^{k+m}} & \text{if } m\geq k+\ell \\
    \aaver{\Omega^{-3k-\ell+m} c_0^{\dag 2k}\ub_\ell,w_{k,m}^{2k+\ell}} & \text{if } m\leq k+\ell
  \end{cases}
\end{equation}
and
\begin{equation}
  b_{ABC} = \begin{cases}
    \aaver{\Omega^{2\ell} c_0^{\dag(m+\ell-k)}\ub_k,w_{m,\ell}^{m+\ell}} & \text{if } m\geq k+\ell \\
    \aaver{\Omega^{2\ell} c_0^{\dag 2\ell}\ub_k,w_{m,\ell}^{k+2\ell}} & \text{if } m\leq k+\ell \,.
  \end{cases}
\end{equation}
By inspecting many examples, we observe that $b_{ABC}$ obeys the symmetry relation:
\begin{equation} \label{eq:sym.beta}
  b_{ABC} = b_{CBA} \,.
\end{equation}

\section{The $6j$-symbols of $U_q(sl_2)$}
\label{app:6j}

Let us consider the heighest-weight modules of $U_q(sl_2)$. We denote the corresponding $6j$-symbols \cite{KR88} as
\begin{equation}
  \tetra aby cdx
\end{equation}
with positive integers $a,b,c,d,x,y$.
In an appropriate gauge (see \cite{KS97}), we have the symmetries
\begin{equation}
  \tetra aby cdx = \tetra bay dcx = \tetra bya dxc 
\end{equation}
and
\begin{equation}
  \tetra aby cdx = \tetra adx cby \,.
\end{equation}
The $F$-symbols are given by:
\begin{equation}
  (F_{abc}^d)_{yx} = \sqrt{[x][y]} \, \tetra aby cdx \,.
\end{equation}
For $d=2$ and $a,c>1$, the $6j$-symbols read
\begin{equation}
  \begin{aligned}
   & \tetra d a {c\pm 1} 2 c {a\pm 1} = \mp (-1)^{\frac{a+c-d}{2}}
   \sqrt{\frac{[\frac{d+a-c}{2}][\frac{d-a+c}{2}]}{[a][c][a\pm 1][c\pm 1]}} \,, \\
   & \tetra d a {c\pm 1} 2 c {a\mp 1} = 
   \sqrt{\frac{[\frac{a+c+d}{2}][\frac{a+c-d}{2}]}{[a][c][a\mp 1][c\pm 1]}} \,.
 \end{aligned}
\end{equation}
This yields
\begin{equation}
  (F_{a2c}^d)_{a\pm 1,c\pm 1} = \mp (-1)^{\frac{a+c-d}{2}}
   \sqrt{\frac{[\frac{d+a-c}{2}][\frac{d-a+c}{2}]}{[a][c]}} \,,
   \qquad  (F_{a2c}^d)_{a\pm 1,c\mp 1} = 
   \sqrt{\frac{[\frac{a+c+d}{2}][\frac{a+c-d}{2}]}{[a][c]}} \,.
\end{equation}
In particular, if we set $d=2$ and $c=a>1$, we have
\begin{equation}
  \begin{aligned}
   & (F_{a2a}^2)_{a\pm 1,a\pm 1} = \frac{\pm (-1)^a}{[a]} \,, 
   \qquad  (F_{a2a}^2)_{a\pm 1,a\mp 1} = 
   \sqrt{\frac{[a+1][a-1]}{[a]^2}} \,,
 \end{aligned}
\end{equation}
which gives the matrix
\begin{equation} \label{eq:F2}
  F_{a2a}^2 = \left(\begin{array}{cc}
    \frac{(-1)^a}{[a]} & \sqrt{\frac{[a+1][a-1]}{[a]^2}} \\
    \sqrt{\frac{[a+1][a-1]}{[a]^2}} & \frac{(-1)^{a+1}}{[a]}
  \end{array}\right)
\end{equation}
in the basis $(a+1,a-1)$.
If $|q|=1$, then $F_{a2a}^2$ is a real unitary matrix.

\end{document}